\documentclass[12pt,aps,prd,preprint,tightenlines,
   showpacs,nofootinbib]{revtex4}
\newcommand{\PRE}[1]{{#1}} 

\usepackage{bm} \usepackage{epsfig}

\newcommand{\mgut}{M_{\text{GUT}}}

\newcommand{\gev}{\text{GeV}}
\newcommand{\tev}{\text{TeV}}
\newcommand{\ifb}{\text{fb}^{-1}}
\newcommand{\fb}{\text{fb}}
\newcommand{\pb}{\text{pb}}

\newcommand{\m}{\text{m}}

\newcommand{\yr}{\text{yr}}

\newcommand{\eqref}[1]{Eq.~(\ref{#1})}

\newcommand{\secref}[1]{Sec.~\ref{sec:#1}}

\newcommand{\figref}[1]{Fig.~\ref{fig:#1}}

\newcommand{\tableref}[1]{Table~\ref{table:#1}}

\begin{document}

\preprint{UCI-TR-2005-40}

\title{
\PRE{\vspace*{1.5in}}
Minimal Supergravity with \boldmath{$m^2_0 < 0$}\\
\PRE{\vspace*{0.3in}} }

\author{Jonathan L.~Feng}
\affiliation{Department of Physics and Astronomy, University of
California, Irvine, CA 92697, USA \PRE{\vspace*{.5in}} }
\author{Arvind Rajaraman}
\affiliation{Department of Physics and Astronomy, University of
California, Irvine, CA 92697, USA \PRE{\vspace*{.5in}} }
\author{Bryan T.~Smith%
\PRE{\vspace*{.2in}} } \affiliation{Department of Physics and
Astronomy, University of California, Irvine, CA 92697, USA
\PRE{\vspace*{.5in}} }

\begin{abstract}
\PRE{\vspace*{.3in}} We extend the parameter space of minimal
supergravity to negative values of $m_0^2$, the universal scalar mass
parameter defined at the grand unified scale.  After evolving to the
weak scale, all scalars can be non-tachyonic with masses consistent
with collider constraints.  This region of parameter space is
typically considered excluded by searches for charged dark matter,
since the lightest standard model superpartner is a charged slepton.
However, if the gravitino is the lightest supersymmetric particle, the
charged slepton decays, and this region is allowed.  This region
provides qualitatively new possibilities for minimal supergravity,
including spectra with light sleptons and very heavy squarks, and
models in which the lightest slepton is the selectron.  We show that
the $m_0^2 < 0$ region is consistent with low energy precision data
and discuss its implications for particle colliders.  These models may
provide signals of supersymmetry in even the first year of operation
at the Large Hadron Collider.
\end{abstract}

\pacs{04.65.+e, 12.60.Jv, 13.85.-t}

\maketitle

\section{Introduction}
\label{sec:introduction}

Supersymmetric models are theoretically motivated extensions of the
standard model (SM) of particle physics that predict both direct and
indirect signals in particle physics experiments.  Most analyses of
supersymmetric models assume the minimal supersymmetric standard model
(MSSM), the supersymmetric extension of the SM that contains the
minimal number of superpartners.  Supersymmetry (SUSY) must also be
broken.  To make phenomenological analyses tractable, a moderately
simple model for soft supersymmetry-breaking terms must be chosen.

By far the most studied supersymmetric model is minimal supergravity
(mSUGRA)~\cite{Chamseddine:1982jx}, which is specified by 6
parameters:
\begin{equation}
m_0^2, M_{1/2}, A_0, \tan\beta, \ \text{sign}(\mu), \ \text{and} \
m_{3/2} \ .
\end{equation}
Here $m_0^2$ is the universal soft scalar mass squared, $M_{1/2}$ is
the universal soft gaugino mass, $A_0$ is the universal soft trilinear
term, $\tan\beta$ is the ratio of the vacuum expectation values of the
up and down type Higgs bosons, $\mu$ is the supersymmetric Higgs mass
parameter, and $m_{3/2}$ is the gravitino mass.  The first three terms
are defined at the grand unified theory (GUT) scale $\mgut \simeq
2.4\times 10^{16}~\gev$, where the gauge couplings unify.  All
superpartner masses and couplings are determined by these 6 parameters
and renormalization group equations (RGEs).  The lightest SM
superpartner is typically either the lighter stau or the lightest
neutralino.

Note that mSUGRA is typically thought to be determined by the first 5
parameters.  When the gravitino is not the lightest supersymmetric
particle (LSP), much of cosmology and all of particle phenomenology is
insensitive to $m_{3/2}$.  However, if the gravitino is the LSP, both
cosmology and particle phenomenology are sensitive to the gravitino
mass, and $m_{3/2}$ is an essential parameter of mSUGRA.

When $R$-parity is conserved, as we assume throughout this study, the
LSP is stable.  Commonly it is (implicitly) assumed that the gravitino
is not the LSP.  In this case, the region of parameter space in which
the stau is the lightest SM superpartner is strongly disfavored, as it
predicts an absolutely stable charged massive particle (CHAMP), which
has not been found~\cite{Smith:1979rz,Dimopoulos:1989hk}.  Results of
mSUGRA analyses are often displayed in the $(m_0^2, M_{1/2})$
plane. Null results from CHAMP searches then exclude from
consideration a thin triangular wedge in this plane with small $m_0^2
> 0$ and, {\em a fortiori}, the entire half plane with $m_0^2 <0$.

As emphasized above, however, this line of reasoning relies on the
assumption that the gravitino is not the LSP.  There are no
theoretical motivations for this assumption --- the gravitino mass is
a free parameter in mSUGRA.  In this, as well as in other scenarios
with high-scale supersymmetry breaking, it is naturally of the same
order of magnitude as other superpartner masses, but it is not
necessarily larger.  In addition, recent work has established that
there are also no phenomenological or cosmological reasons to exclude
the gravitino LSP scenario~\cite{Feng:2003xh,Ellis:2003dn,%
Feng:2004zu,Wang:2004ib,Ellis:2004bx,Roszkowski:2004jd,%
Brandenburg:2005he,Cerdeno:2005eu}.  In fact, high-scale supersymmetry
breaking with a gravitino LSP has a number of novel implications and
virtues.  For example, if the gravitino is the LSP and the
next-lightest supersymmetric particle (NLSP) is charged, the signal of
SUSY at colliders will be metastable charged particles. Such particles
have spectacular signatures~\cite{Drees:1990yw,Goity:1993ih,%
Nisati:1997gb,Feng:1997zr}.  They also make possible the investigation
of gravitational interactions and the quantitative verification of
supergravity in high energy physics
experiments~\cite{Buchmuller:2004rq,Feng:2004gn,Hamaguchi:2004df,%
Feng:2004yi,DeRoeck:2005bw}.  Cosmologically, gravitinos produced
through decays naturally have the correct relic density to be
superweakly-interacting massive particle (superWIMP) dark matter. For
some parameters, gravitino dark matter produced in this way has
features usually associated with warm dark matter and may resolve
controversial discrepancies in halo profiles and the formation of
small scale structure~\cite{Kaplinghat:2005sy,Cembranos:2005us,%
Jedamzik:2005sx,Sigurdson:2003vy,Profumo:2004qt}.  Last, the late
decays that produce gravitino dark matter also release electromagnetic
and hadronic energy, with (possibly felicitous) implications for Big
Bang nucleosynthesis~\cite{Feng:2003xh,Feng:2004zu,Roszkowski:2004jd,%
Cerdeno:2005eu,Jedamzik:2004er,Kawasaki:2004qu,Ellis:2005ii,%
Jedamzik:2005dh} and the cosmic microwave
background~\cite{Feng:2003xh,Feng:2004zu,Lamon:2005jc}.

In this paper we consider the possibility of mSUGRA with a gravitino
LSP and $m_0^2 < 0$. We define
\begin{equation}
m_0 \equiv \text{sign}(m^2_0)\sqrt{|m^2_0|} \ . \label{convention}
\end{equation}
We show that the region with $m_0 < 0$ contains models consistent with
all collider limits.  We also consider precision measurements,
analyzing the anomalous magnetic moment of the muon $a_{\mu}$, $B(b
\to s\gamma)$, and $B_s^0 \to \mu^+ \mu^-$.  We find that the current
discrepancy in $a_{\mu}$ between experiment and the SM prediction may
be resolved for $m_0 <0$ without disrupting the agreement for $B(b \to
s\gamma)$, and near future probes of $B_s^0 \to \mu^+ \mu^-$ will have
significant reach in $m_0 < 0$ model space.  Precision data do not
currently favor one sign of $m_0$ over the other.

The simple modification of taking $m_0<0$ therefore ``doubles'' the
viable mSUGRA parameter space and leads to qualitatively new
possibilities.  For example, in some regions of parameter space, the
NLSP is not the stau, but the selectron.  This overturns the common
lore that Yukawa couplings in RGEs lower soft masses; when some
scalars are tachyonic in part of the RG evolution, Yukawa terms may
{\em increase} scalar masses.  We also find that light charged
sleptons can be produced for any value of $M_{1/2}$.  This produces
spectra where the charged sleptons have masses around $100~\gev$, but
all other superpartner masses are above 1 TeV, with squark and gluino
masses around $3-4~\tev$.  Such spectra are not found for $m_0 > 0$
and have novel implications for the Large Hadron Collider (LHC) and
International Linear Collider (ILC).  Although all of these features
may be found in general MSSM models, it is striking that we find them
here in a framework with universal scalar and gaugino masses,
motivated as these features are by simplicity, the SUSY flavor and CP
problems, and gauge coupling unification.

Cosmologically, these models differ from conventional models with $m_0
>0$ in several ways.  As noted above, the $m_0 <0$ models have
superWIMP dark matter, as opposed to the conventional neutralino WIMP
dark matter, with the implications mentioned above for Big Bang
nucleosynthesis, the cosmic microwave background, and structure
formation. In addition, there are possibly novel implications for
vacuum stability~\cite{Cerdeno:2005eu} and gauge symmetry breaking at
high temperatures.  We defer discussion of cosmological
issues~\cite{next}, and focus here on implications for particle
physics.

This study is organized as follows.  In \secref{rge} we show that,
even given $m_0<0$ at the GUT scale, all superpartner masses, when
evolved to the weak scale, can have values consistent with current
experimental bounds. We delineate the allowed regions and determine
which regions of parameter space have stau and selectron NLSPs.  The
resulting superpartner masses in the $m_0<0$ region are discussed in
\secref{mass}. Low-energy observables are analyzed in
\secref{constraints}.  In \secref{collider}, we show two
representative superpartner spectra and briefly discuss the
implications for the LHC and ILC.  In \secref{conclusions}, we
conclude and indicate interesting avenues for further investigation.

\section{Regions of mSUGRA Parameter Space for
\boldmath{$\lowercase{m}_0 < 0$}} \label{sec:rge}

In this section, we determine the allowed regions of mSUGRA parameter
space with $m_0 <0$, and further classify the allowed parameter space
according to what particles are the LSP and NLSP, since these play a
large role in determining experimental signatures.

For $m_0<0$, an immediate worry is that scalar masses will remain
tachyonic even after RG evolution to the weak scale.  As usual, gauge
interactions raise the soft masses, and so the most problematic
scalars are the right-handed sleptons, since these have only
hypercharge interactions.  The RGEs for conventional mSUGRA have been
studied in great detail.  A well-known approximate relation for the
weak-scale right-handed selectron mass in terms of GUT-scale
parameters is~\cite{Polonsky:2001pn}
\begin{equation}
m_{\tilde{e}_R}^2 = m_0^2 + 0.15 \, M_{1/2}^2 \ . \label{ermass}
\end{equation}
This remains valid for $m_0<0$.  {}From this, we see that negative
$m_0^2$ can always be compensated by large $M_{1/2}$ to make the
right-handed sleptons, and with them the entire superpartner
spectrum, non-tachyonic.

\begin{figure}
\resizebox{6.0in}{!}{
\includegraphics{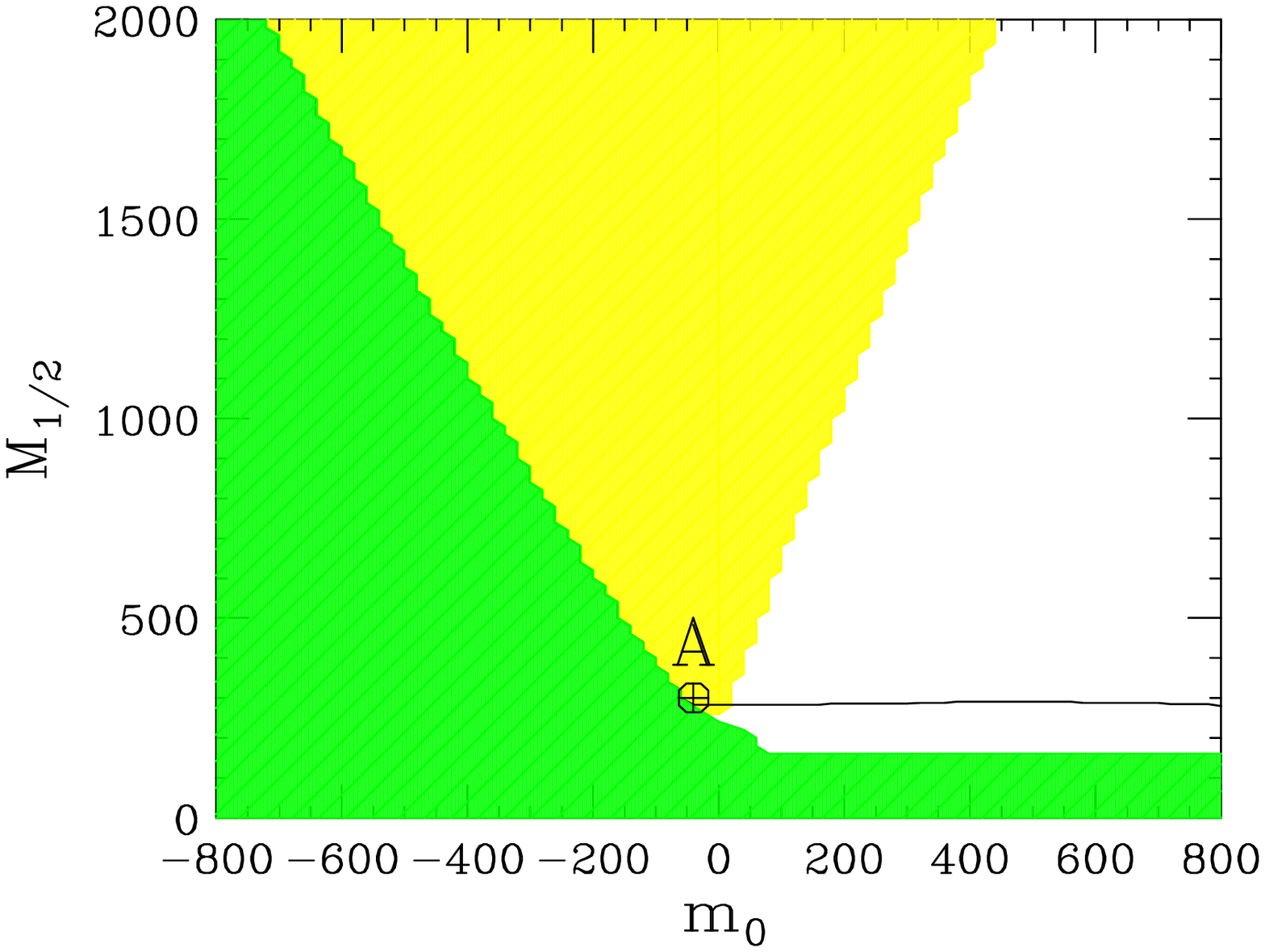} \qquad
\includegraphics{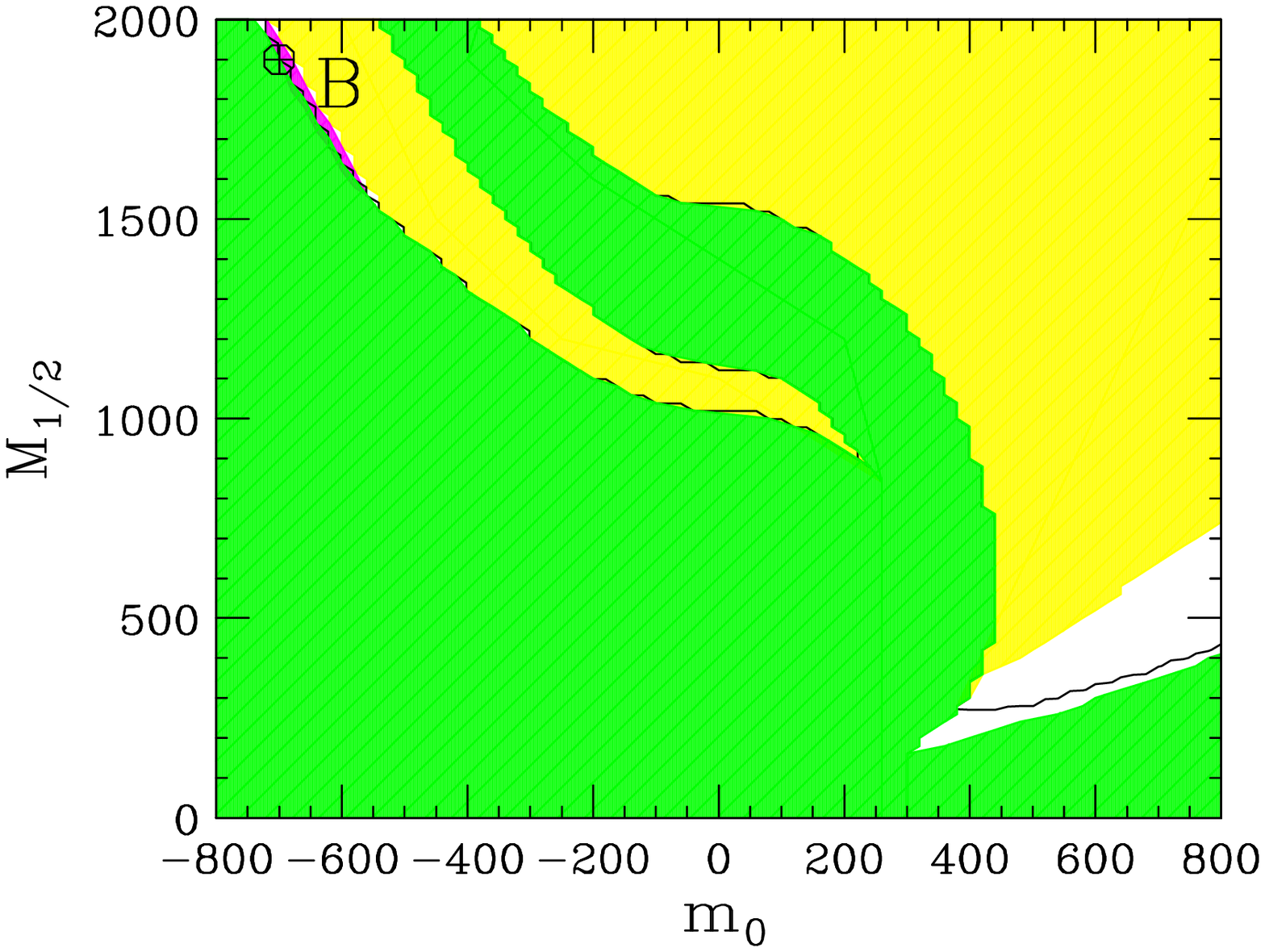}
} 
\caption{Regions of the $(m_0, M_{1/2})$ plane, extended to $m_0 < 0$,
for $A_0 = 0$, $\mu > 0$, and $\tan\beta = 10$ (left) and $\tan\beta =
60$ (right).  The green (medium shaded) region is experimentally
excluded, and the unshaded region is the conventional neutralino
(N)LSP region.  In the remaining regions, the gravitino is the LSP: in
the yellow (light shaded) region, the stau is the NLSP, and in the
thin magenta (dark shaded) region of the $\tan\beta = 60$ plot, the
selectron is the NLSP.  The present Higgs mass bound $m_h >
114.1~\gev$ excludes regions below the solid contours.  The symbols
$\oplus$ mark the location of benchmark Models A and B; their RG
evolution is shown in \figref{rgelepton}. \label{fig:lsp} }
\end{figure}

Allowed regions of the $(m_0, M_{1/2})$ plane are shown for two values
of $\tan\beta$ in \figref{lsp}.  The SUSY spectra have been calculated
with the software package ISAJET v7.71~\cite{Paige:2003mg}, modified
to accommodate $m_0<0$. ISAJET includes 2-loop RGEs and 1-loop
corrections to superpartner masses, and we choose a top quark mass of
175 GeV.  The green (medium shaded) region is excluded.  For $m_0>0$
the boundary is determined by the LEP chargino mass limit
$m_{\tilde{\chi}^{\pm}} > 103.5~\gev$~\cite{LEPchargino}.  For
$m_0<0$, it is essentially determined by null searches for long-lived
charged sleptons at LEP, leading to limits $m_{\tilde{l}_R} >
99~\gev$~\cite{LEPstableslepton}. For $\tan\beta = 10$, the border for
$m_0<0$ follows to a reasonable approximation the tachyonic slepton
line $m_0 = - 2.6 M_{1/2}$ one can derive from \eqref{ermass}.  For
$\tan\beta = 60$, the excluded region has a more complicated shape
because large 1-loop corrections play an important role, as discussed
below.

The allowed regions may be further divided according to what
particles are the LSP and NLSP.  The unshaded regions of \figref{lsp}
are the conventional regions in which the lightest SM superpartner is
the lightest neutralino.  It is either the LSP or, if the gravitino
is the LSP, the NLSP.

In the rest of the allowed regions shown, the gravitino must be the
LSP to avoid having charged dark matter, and the NLSP is a charged
slepton.  To determine which charged slepton is the NLSP, consider
the RGEs for their soft mass parameters.  At 1-loop, these are
\begin{eqnarray}
\frac{d m^2_{\tilde{e}_{R}}}{dt} &=& \frac{2}{16\pi^{2}}\left[
-\frac{12}{5}g^2_1 M^2_1 \right]
\\
\frac{d m^2_{\tilde{\tau}_{R}}}{dt} &=& \frac{2}{16\pi^2} \left[
-\frac{12}{5}g^2_1 M^2_1 + 2\lambda^2_{\tau}
\left(m^2_{\tilde{\tau}_{L}} + m^2_{\tilde{\tau}_{R}} + m^2_{H_{d}} +
A^2_{\tilde{\tau}} \right) \right] \ , \label{tauRrge}
\end{eqnarray}
where $t=\ln\left( Q^2 / \mgut^2 \right)$.  As is well-known, when
all mass parameters are non-tachyonic, Yukawa interactions lower soft
masses, leading to the common lore that, given universal scalar
boundary conditions, the lightest slepton is always the stau.  In the
present case, however, $m_0 < 0$, and so in evolving from the GUT
scale, selectron masses initially rise slower than stau masses.  Of
course, for the spectrum to be viable, all physical scalar masses
must eventually become non-tachyonic, and so will exert the
conventional effect of Yukawa couplings as one approaches the weak
scale.  (The Higgs scalar mass parameters may remain negative.) The
competition between the new and the conventional effects determines
whether the NLSP is the selectron or the stau.

\begin{figure}
\resizebox{6.0in}{!}{
\includegraphics{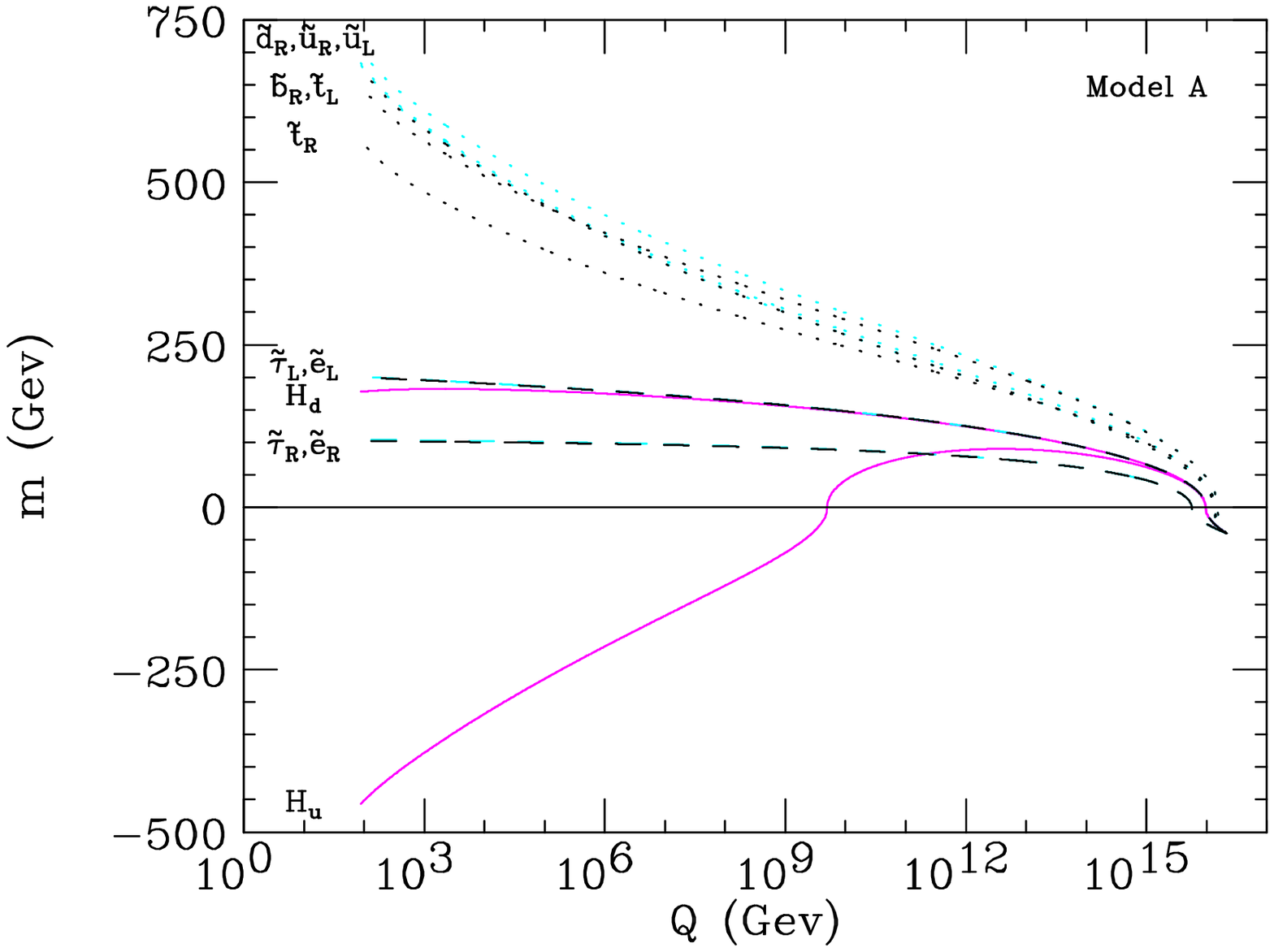} \qquad
\includegraphics{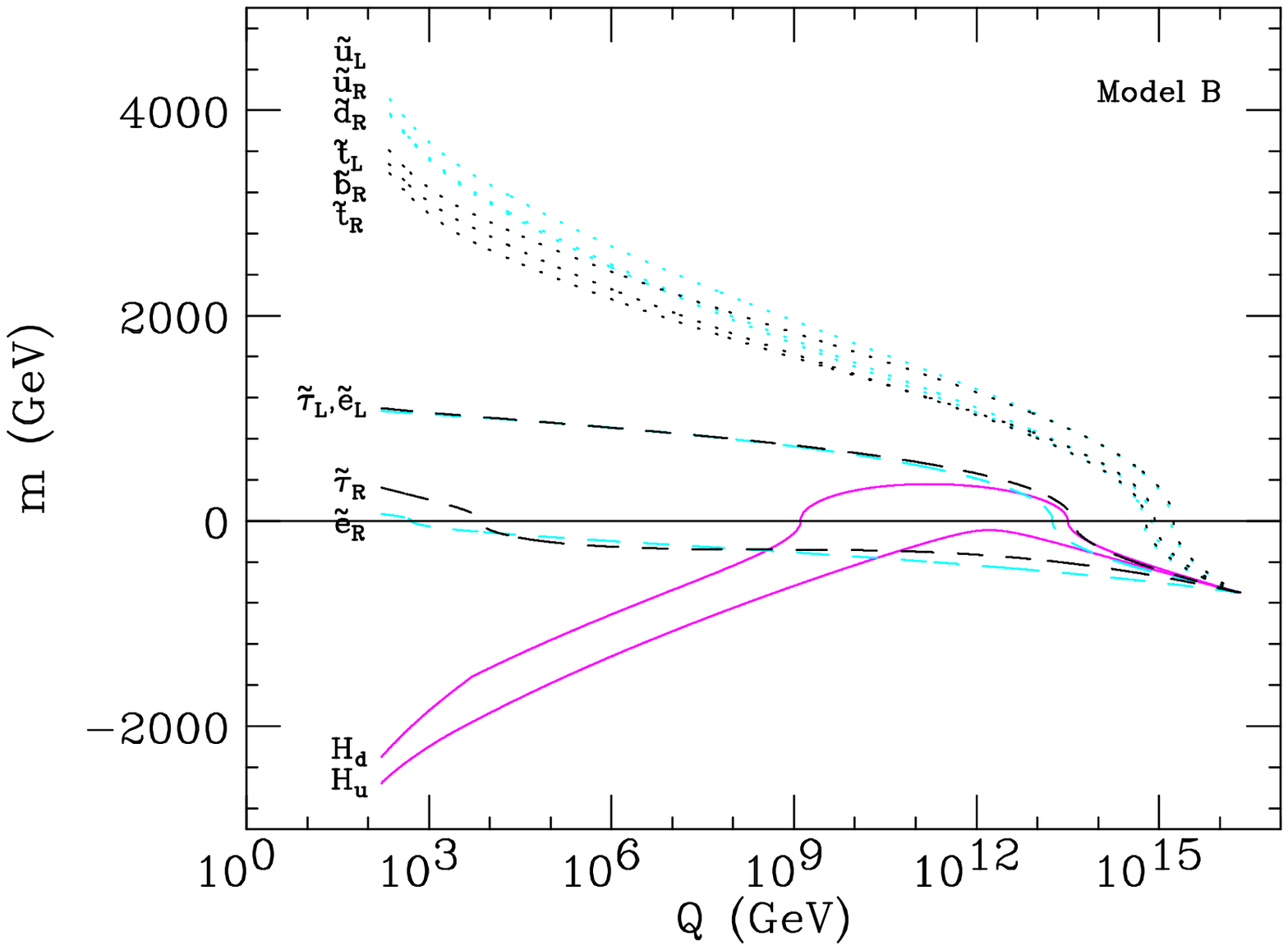}
} 
\caption{The RG evolution of soft scalar masses in Model A, the stau
NLSP point with $m_0 = -40~\gev$, $M_{1/2} = 300~\gev$, $\tan\beta =
10$ (left) and in Model B, the selectron NLSP point with $m_0 =
-700~\gev$, $M_{1/2} = 1900~\gev$, $\tan\beta$ = 60 (right).  In both
cases, $A_0 = 0$ and $\mu > 0$.
\label{fig:rgelepton} }
\end{figure}

These effects are shown in \figref{rgelepton} for the two benchmark
models highlighted in \figref{lsp}.  In the left panel, all scalar
masses of Model A quickly become positive as they evolve from the GUT
scale, and so the stau becomes the NLSP, as usual.  In the right
panel, however, the scalar masses of Model B are negative for much of
the RG evolution, and $m_{H_d}^2$ becomes negative, leading to an
inverted flavor spectrum with a selectron NLSP.

In \figref{lsp}, the $\tilde{\tau}_1$ is the NLSP in the yellow (light
shaded) region, and $\tilde{e}_R$ is the NLSP in the magenta (dark
shaded) region.  The current experimental limits force sleptons to be
not just non-tachyonic, but significantly so, and so the viable
selectron NLSP region is reduced to a thin sliver near the upper,
left-hand corner in the $\tan\beta = 60$ plot.  Its exact location is
therefore rather sensitive to small corrections and depends on the
implementation of RG evolution and loop-level corrections to the
superpartner mass spectrum. The mere possibility that universal
slepton masses can lead to selectrons lighter than staus, however, is
a robust physics effect; it is novel and never realized in
conventional mSUGRA.  The selectron NLSP region may be much larger in
even slightly more general models.  For example, motivated by SO(10)
unification, one may consider models in which the matter scalar masses
are determined by the parameter $m_0^2$, but the Higgs scalar masses
are unified at a different GUT-scale parameter $m_H^2$.  In these
non-universal Higgs mass models~\cite{Baer:2005bu}, by choosing $m_H^2
< m_0^2 < 0$, the stau mass will receive large positive contributions
from $m_{H_d}^2$ that are absent for the selectron, as can be seen in
\eqref{tauRrge}, and the selectron NLSP region will be much larger.
Last, we note that in the mSUGRA models considered here, the
possibility of a non-tachyonic sneutrino (N)LSP also exists at low
$M_{1/2}$, but this lies entirely in the excluded region.

As noted above, the excluded region can have a rather complicated
shape.  For $\tan\beta=60$, for example, as can be seen in the
right-hand panel of \figref{lsp}, the excluded region has an
interesting shape.  This results from the remarkable fact that the
light charged slepton masses do not increase monotonically as one
increases $M_{1/2}$ for constant $m_0$.  In fact, this is not peculiar
to $m_0<0$, as it occurs even for {\em positive} constant $m_0$.

\begin{figure}
\resizebox{3.0in}{!}{
\includegraphics{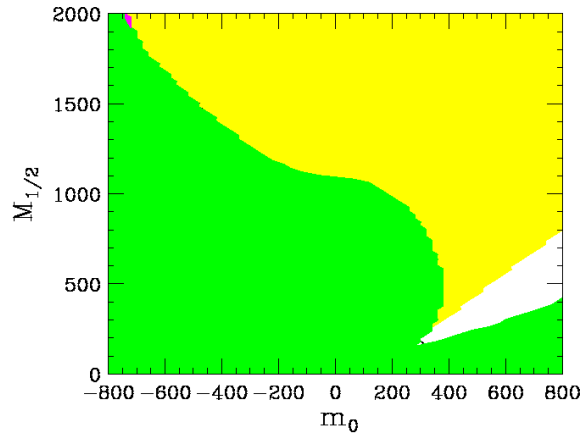}
} \caption{ Regions of the $(m_0, M_{1/2})$ plane, extended to $m_0 <
0$, for $A_0 = 0$, $\mu > 0$, and $\tan\beta = 60$ as in the
right-hand panel of \protect\figref{lsp}, but with 1-loop corrections
to sparticle masses neglected. \label{fig:lspnoloop} }
\end{figure}

This behavior results from 1-loop corrections present in the slepton
mass matrix
\begin{equation}
\left(
\begin{array}{cc}
M^2_{LL}+\delta M^2_{LL} & \quad M^2_{LR}+\delta M^2_{LR} \\
M^2_{RL}+\delta M^2_{RL} & \quad M^2_{RR}+\delta M^2_{RR}
\end{array}
\right) \ , \label{massmatrix}
\end{equation}
where $M^2$ are tree-level contributions and $\delta M^2$ are 1-loop
corrections.  If 1-loop corrections are neglected, the slepton mass
monotonically increases for increasing $m_0$ and fixed $M_{1/2}$ (or
vice versa), as illustrated in \figref{lspnoloop}.  However, for
large $\tan\beta$ the loop corrections have the proper sign and
magnitude to lower the mass eigenvalues of the stau below the
experimental bounds in part of the parameter space.  Of course, the
1-loop corrections are physical, and we include them in all plots and
results below.

To conclude this section, the impression that neutralinos are the
lightest SM superpartners in most of mSUGRA parameter space is
artificial: it follows only if one requires $m_0>0$.  Allowing
$m_0<0$ extends the viable region of mSUGRA parameter space
significantly and shows that staus are the lightest SM superpartners
in much of mSUGRA parameter space.  In addition, allowing $m_0<0$
leads to other new phenomena, such as the possibility that the
selectron is the lightest SM superpartner.  In the next section, we
explore the implications of $m_0<0$ for the sparticle spectrum more
fully.

\section{SUSY Mass Spectra for \boldmath{$\lowercase{m}_0 < 0$}}
\label{sec:mass}

\begin{figure}
\resizebox{6.0in}{!}{
\includegraphics{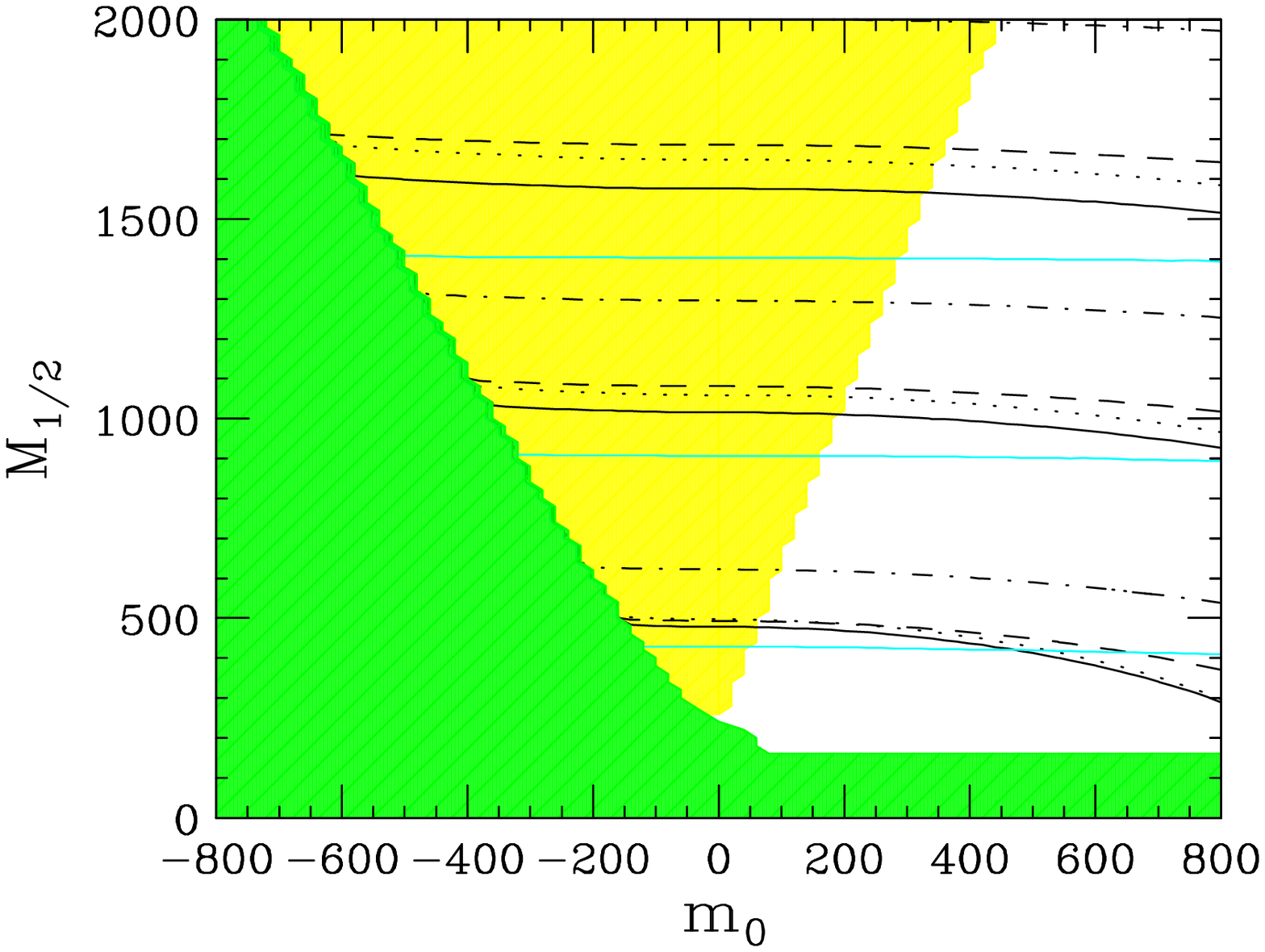} \qquad
\includegraphics{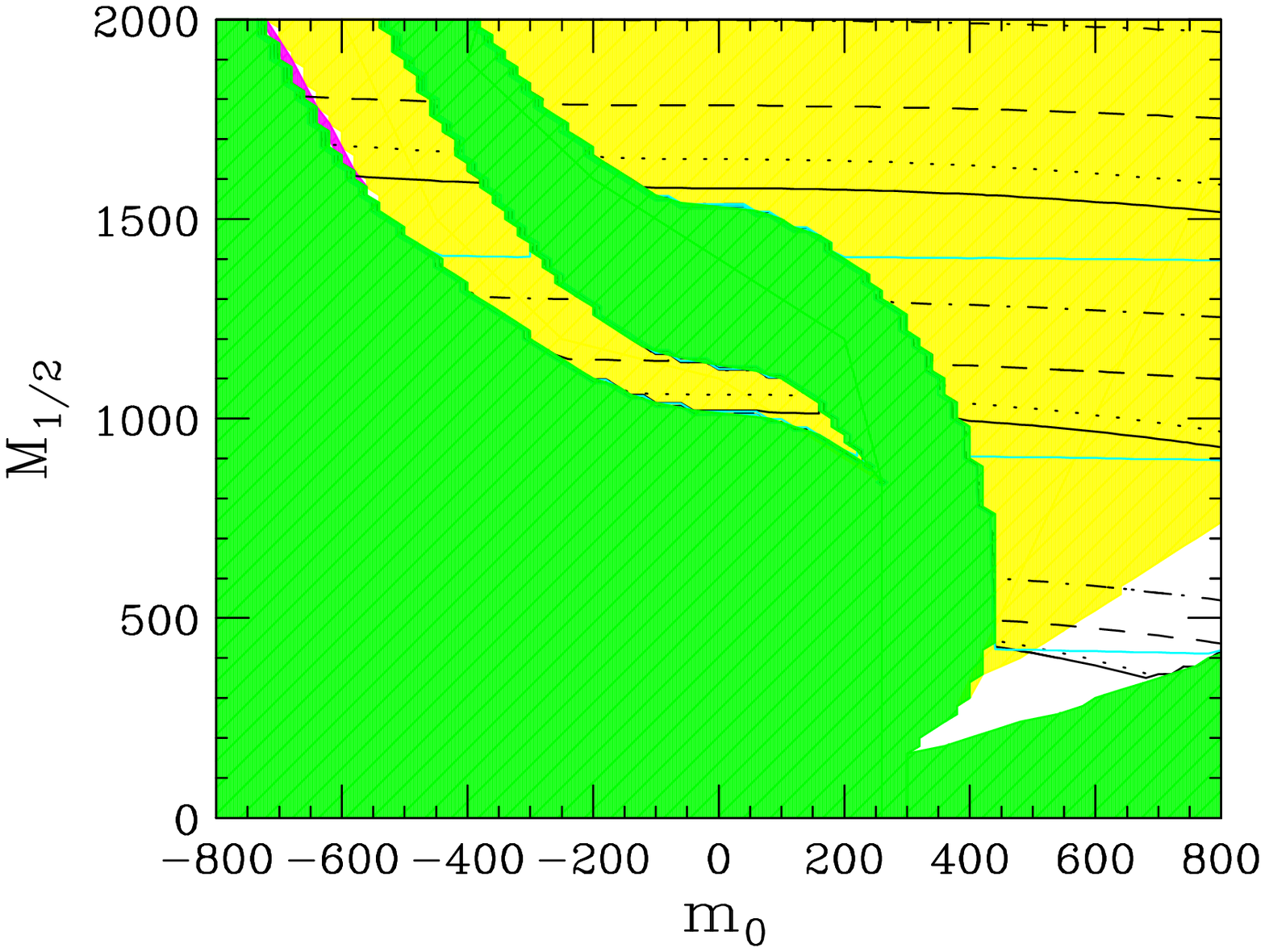}
} \resizebox{6.0in}{!}{
\includegraphics{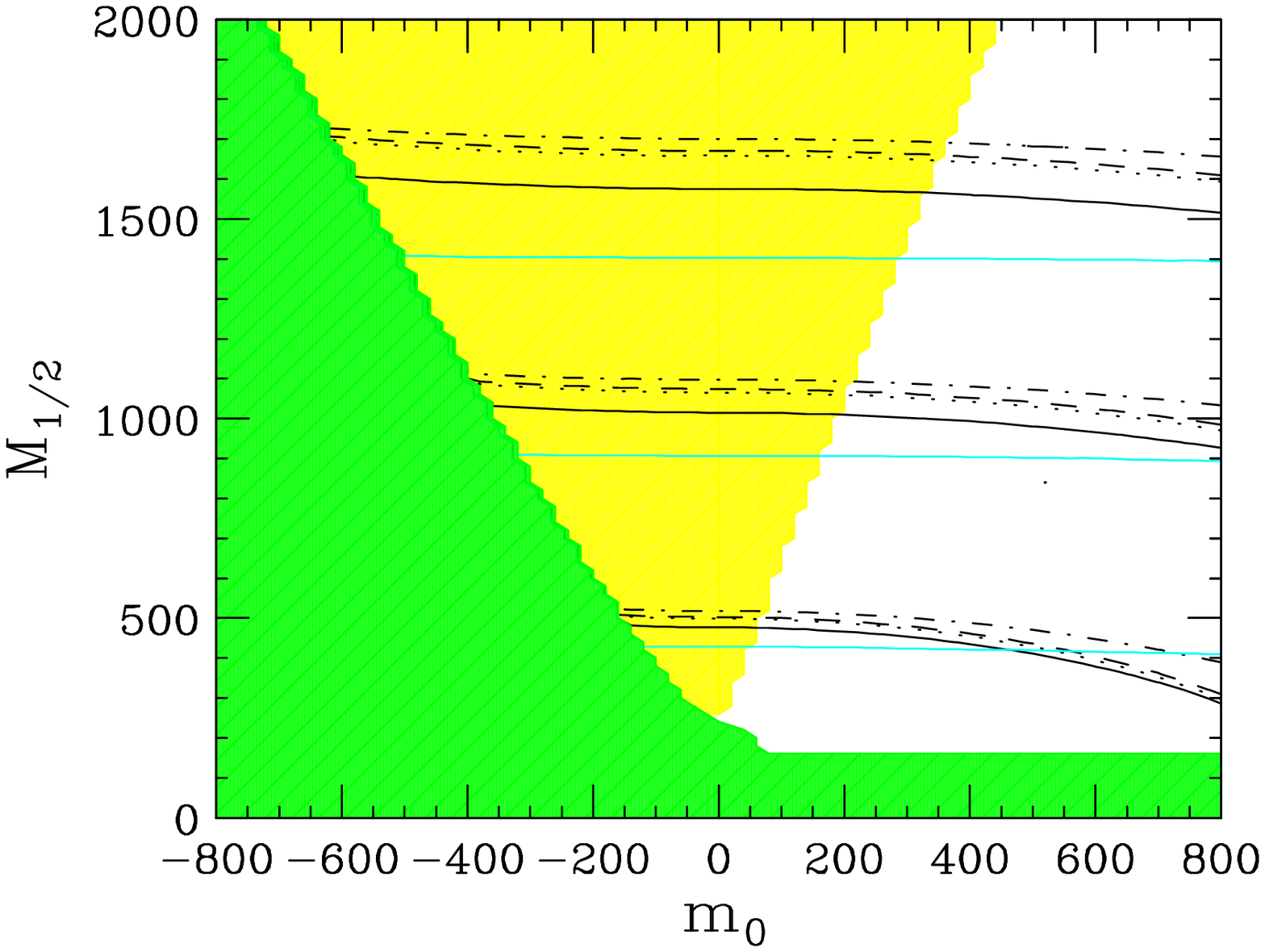} \qquad
\includegraphics{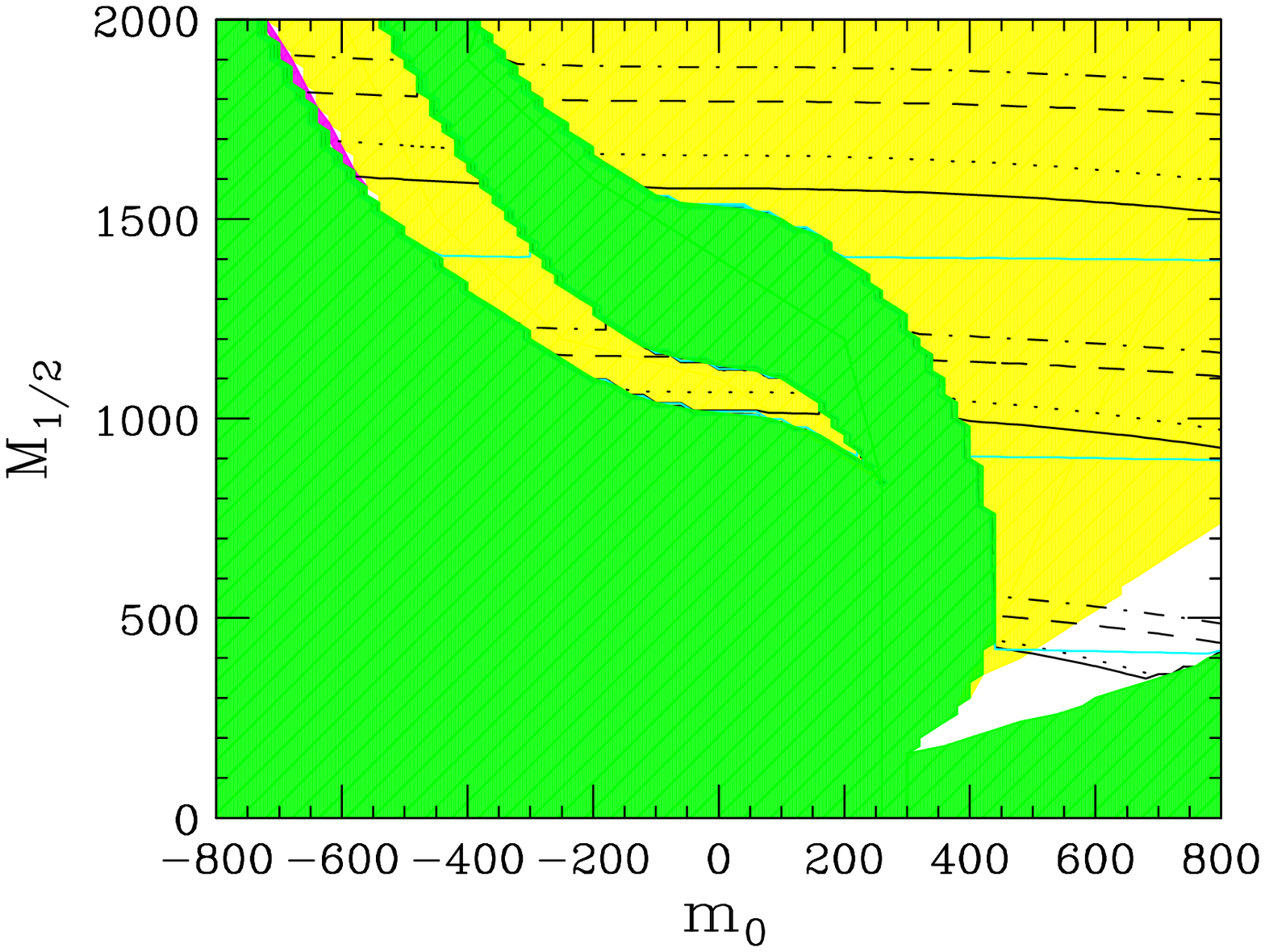}
} 
\caption{Squark masses for mSUGRA extended to $m_0<0$ for $A_0 = 0$,
$\mu > 0$, and $\tan\beta = 10$~(left) and $\tan\beta =
60$~(right). The up-type squarks~(top) are $\tilde{u}_{L}$~(solid
black), $\tilde{u}_{R}$~(dotted black), $\tilde{t}_1$~(dot dashed
black), and $\tilde{t}_2$~(dashed black). Similarly the down-type
squarks~(bottom) are $\tilde{d}_{L}$~(solid black),
$\tilde{d}_{R}$~(dotted black), $\tilde{b}_1$~(dot dashed black), and
$\tilde{b}_2$~(dashed black). The gluino mass~(cyan, solid light) is
presented on all 4 plots.  The contours are for masses $1~\tev$,
$2~\tev$, and $3~\tev$ from bottom to top. \label{fig:quark}}
\end{figure}

The squark and gluino masses are presented in the $(m_0, M_{1/2})$
plane in \figref{quark}.  In the allowed region with $m_0<0$, they
are relatively insensitive to $m_0$, since they are dominated by the
RG contributions of the gaugino masses. The left-handed down and up
squarks are approximately degenerate, as are the right-handed up and
down squarks.  The gluino is always heavier than all squarks in the
$m_0<0$ allowed region.

\begin{figure}
\resizebox{6.0in}{!}{
\includegraphics{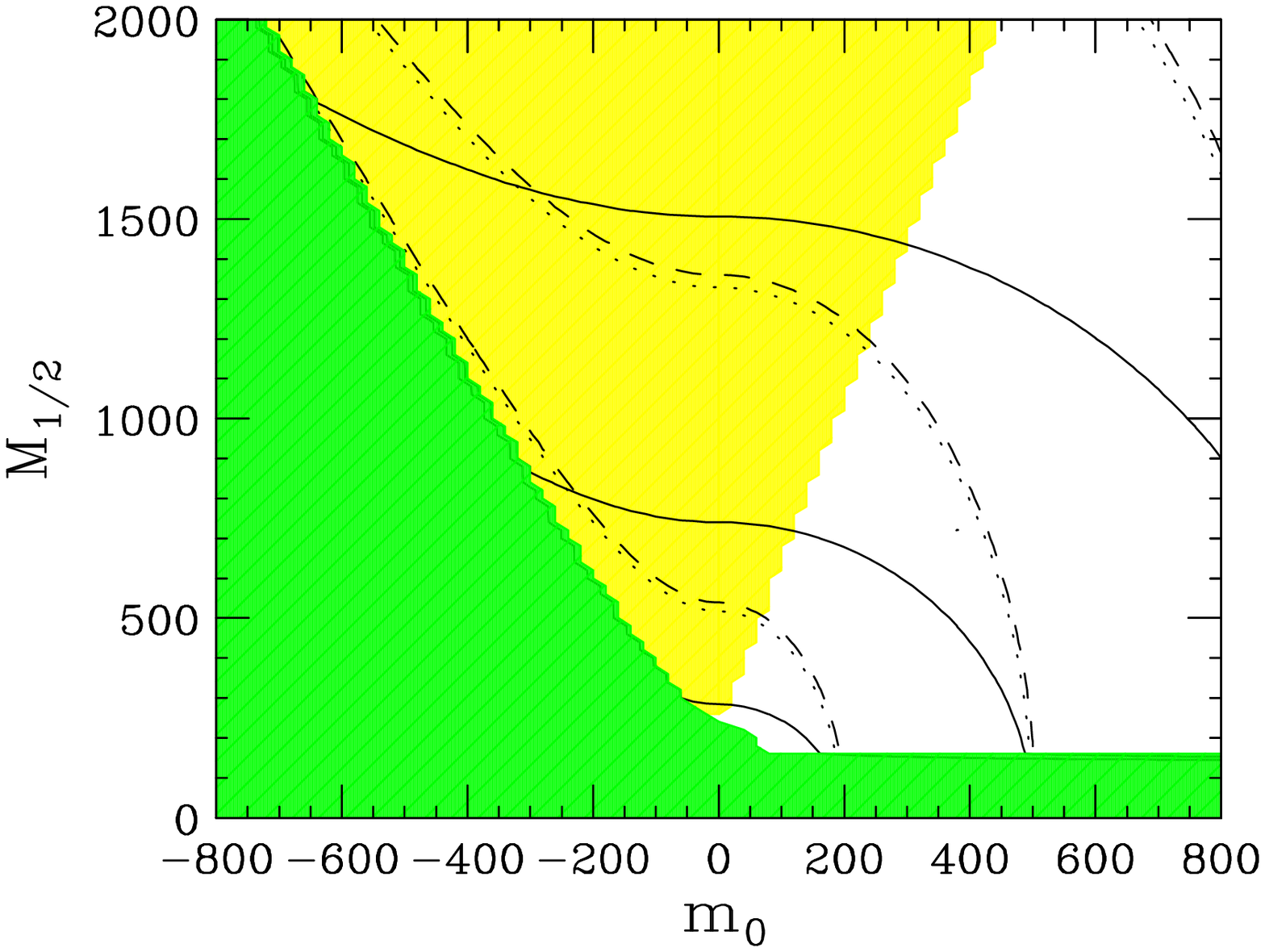} \qquad
\includegraphics{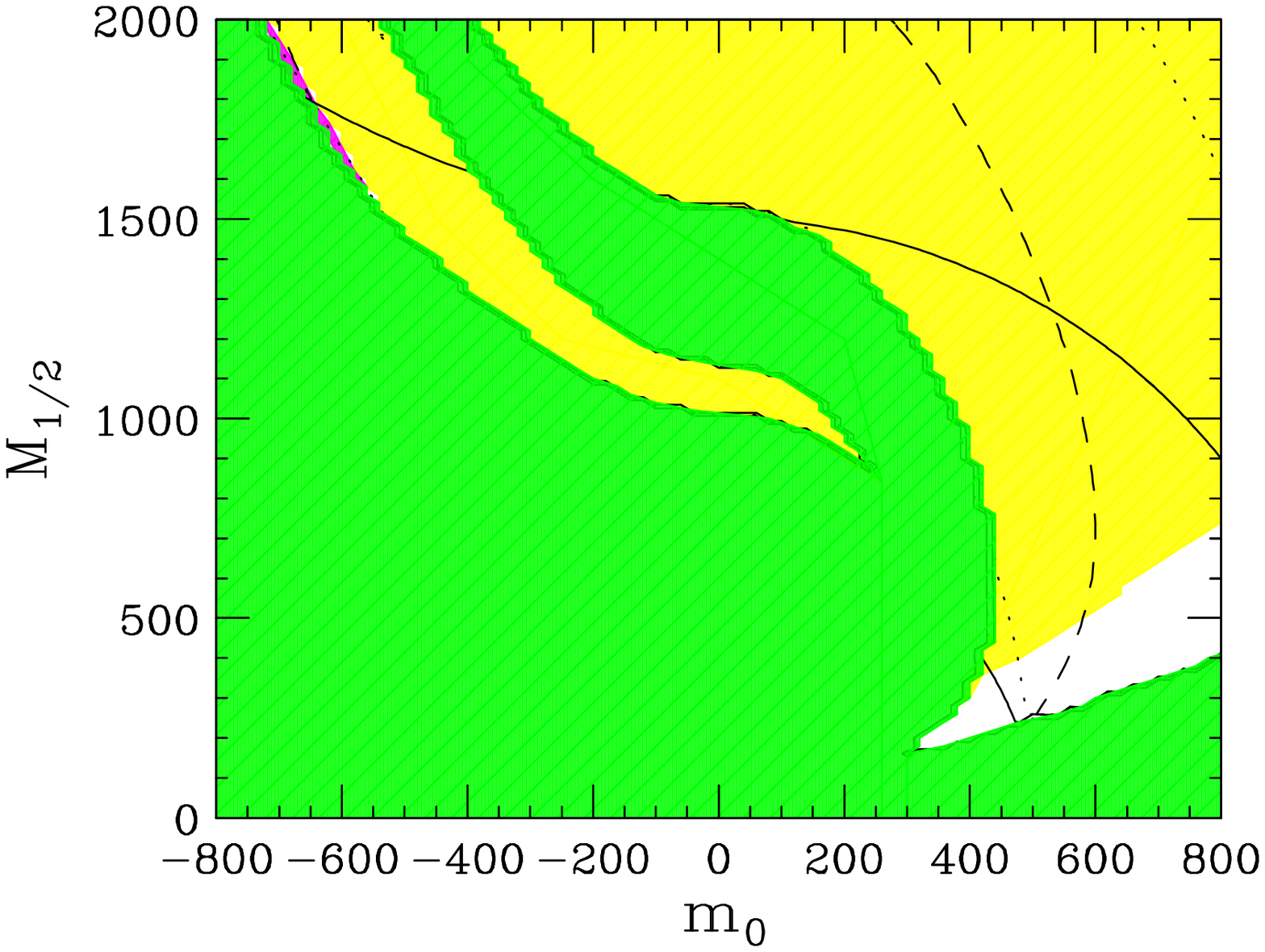}
} \caption{Slepton masses for mSUGRA extended to $m_0<0$ for $A_0 =
0$, $\mu > 0$, and $\tan\beta = 10$~(left) and $\tan\beta =
60$~(right). The contours are for $\tilde{\tau}_1$~(dashed),
$\tilde{e}_{L}$~(solid) and $\tilde{e}_{R}$~(dotted).  The contours
are for masses $200~\gev$, $500~\gev$, and $1~\tev$ from bottom left
to top right, with the exception that in the right panel, because the
slepton masses drop and then rise again for fixed $M_{1/2}$ and
increasing $m_0$, there are two sets of 200 GeV contours for both
$\tilde{e}_R$ and $\tilde{\tau}_1$.  For both of these particles, the
leftmost 200 GeV contour is barely visible in the selectron NLSP
region.  Throughout the parameter space, the sneutrinos and
$\tilde{\tau}_2$ are almost degenerate with the $\tilde{e}_{L}$.
\label{fig:lepton}}
\end{figure}

Slepton masses are shown in \figref{lepton}.  In contrast to squark
masses, slepton masses are extremely sensitive to $m_0$ in the allowed
$m_0<0$ region.  Contours of constant slepton mass switch from concave
down for $m_0 > 0$ to concave up for $m_0 < 0$.  As a result, light
sleptons near their experimental limit may be found for any value of
$M_{1/2}$.  This is true even though the masses of all other
superpartners becomes large for large $M_{1/2}$, as may be seen in
Model B, the selectron NLSP model shown in the right panel of
\figref{rgelepton}.

\begin{figure}
\resizebox{6.0in}{!}{
\includegraphics{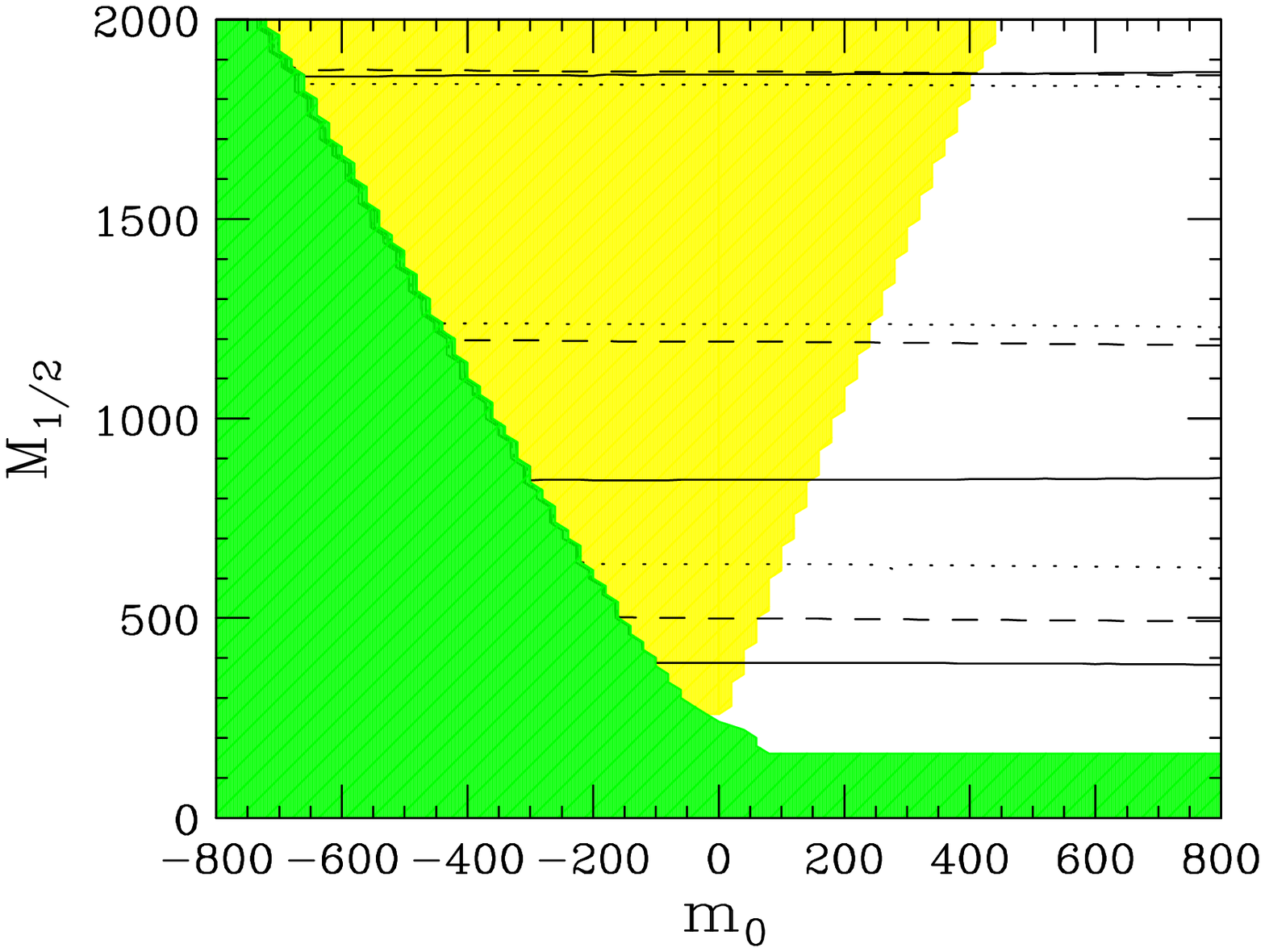} \qquad
\includegraphics{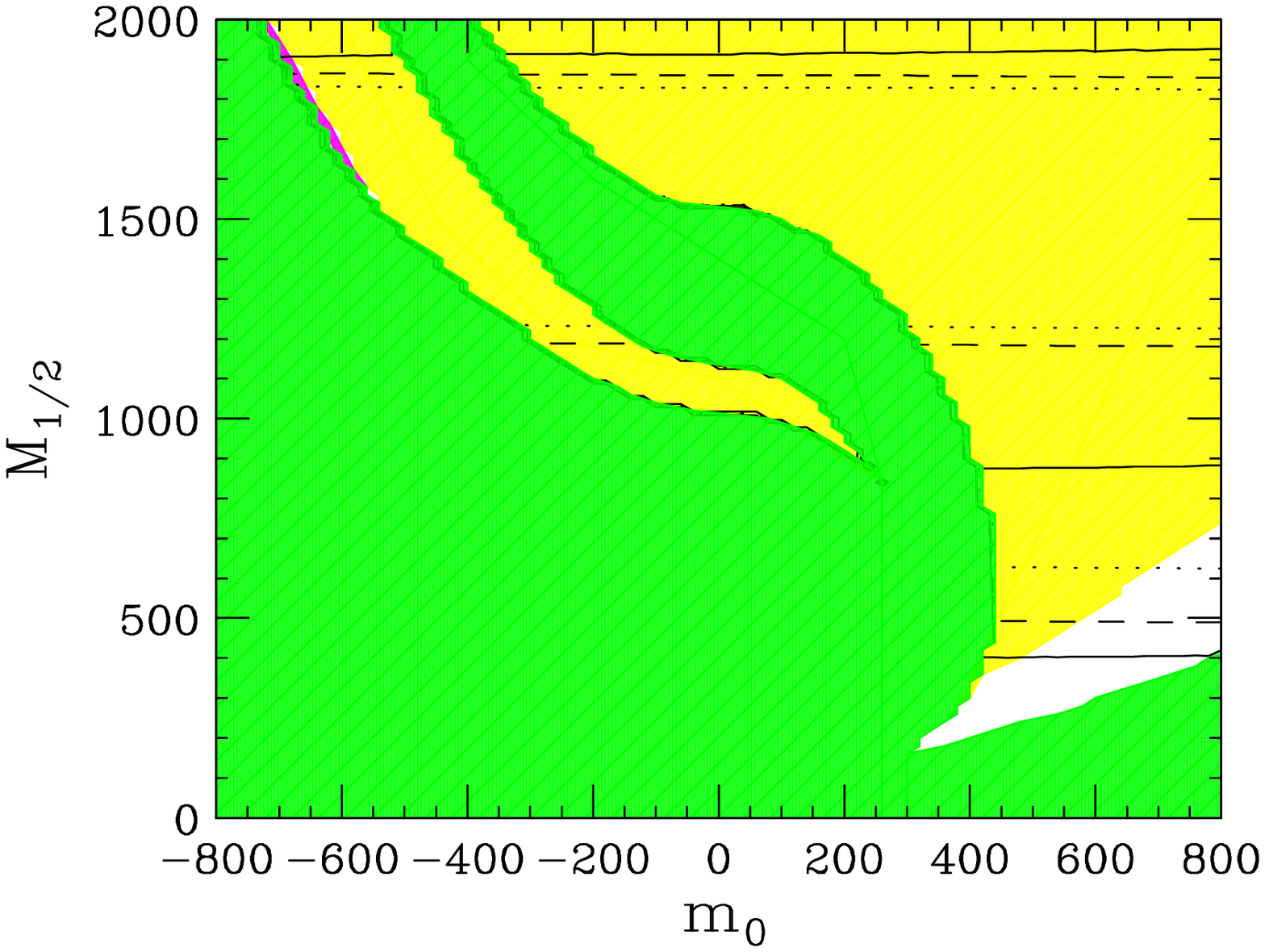}
} 
\caption{Neutralino and chargino masses for mSUGRA extended to $m_0<0$
for $A_0 = 0$, $\mu > 0$, and $\tan\beta = 10$~(left) and $\tan\beta =
60$~(right).  The lightest neutralino $\tilde{\chi}^0_1$~(dashed) has
values from bottom to top of $200~\gev$, $500~\gev$, and $800~\gev$.
The lightest chargino $\tilde{\chi}^+_1$~(dotted) has values from
bottom to top of $500~\gev$, $1~\tev$, and $1.5~\tev$, while
$\mu$~(solid) has values $500~\gev$, $1~\tev$, and $2~\tev$ from
bottom to top.
\label{fig:neutralino}}
\end{figure}

In \figref{neutralino}, we present contours for neutralino and
chargino masses and for the Higgsino mass parameter $\mu$. In the
$m_0 < 0$ region, $|\mu|$ is always much larger than the electroweak
gaugino masses $M_1$ and $M_2$.  As a result, the lighter chargino
and lighter two neutralinos are nearly pure gauginos, with
$\tilde{\chi}^0_1 \approx \tilde{B}$, $\tilde{\chi}^0_2 \approx
\tilde{W}^0$, $\tilde{\chi}^\pm_1 \approx \tilde{W}^{\pm}$, and
$m_{\tilde{\chi}^0_1} \approx M_1$ and $m_{\tilde{\chi}^0_2} \approx
m_{\tilde{\chi}^\pm_1} \approx M_2 \approx 2 M_1$.

\begin{figure}
\resizebox{6.0in}{!}{
\includegraphics{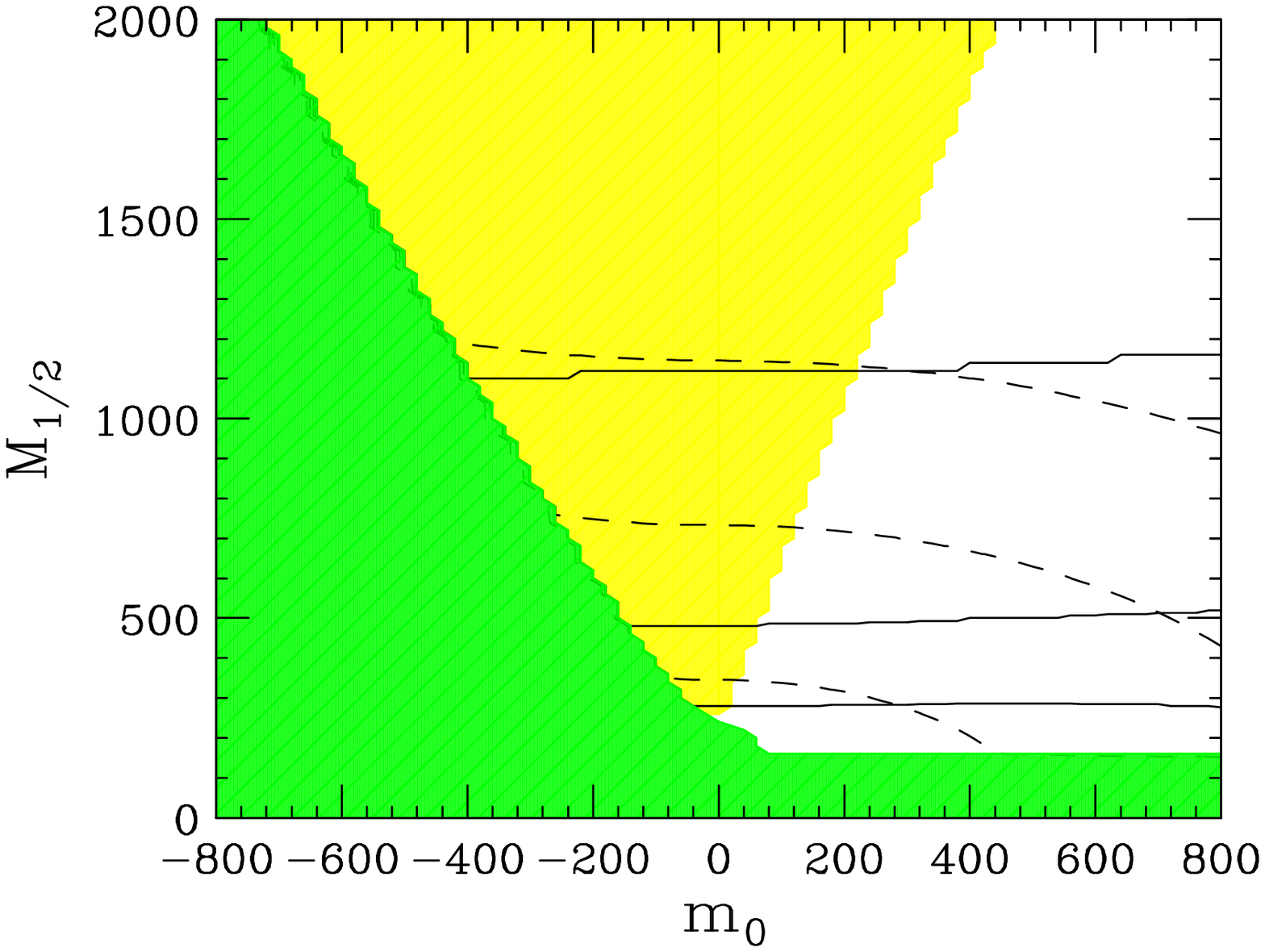} \qquad
\includegraphics{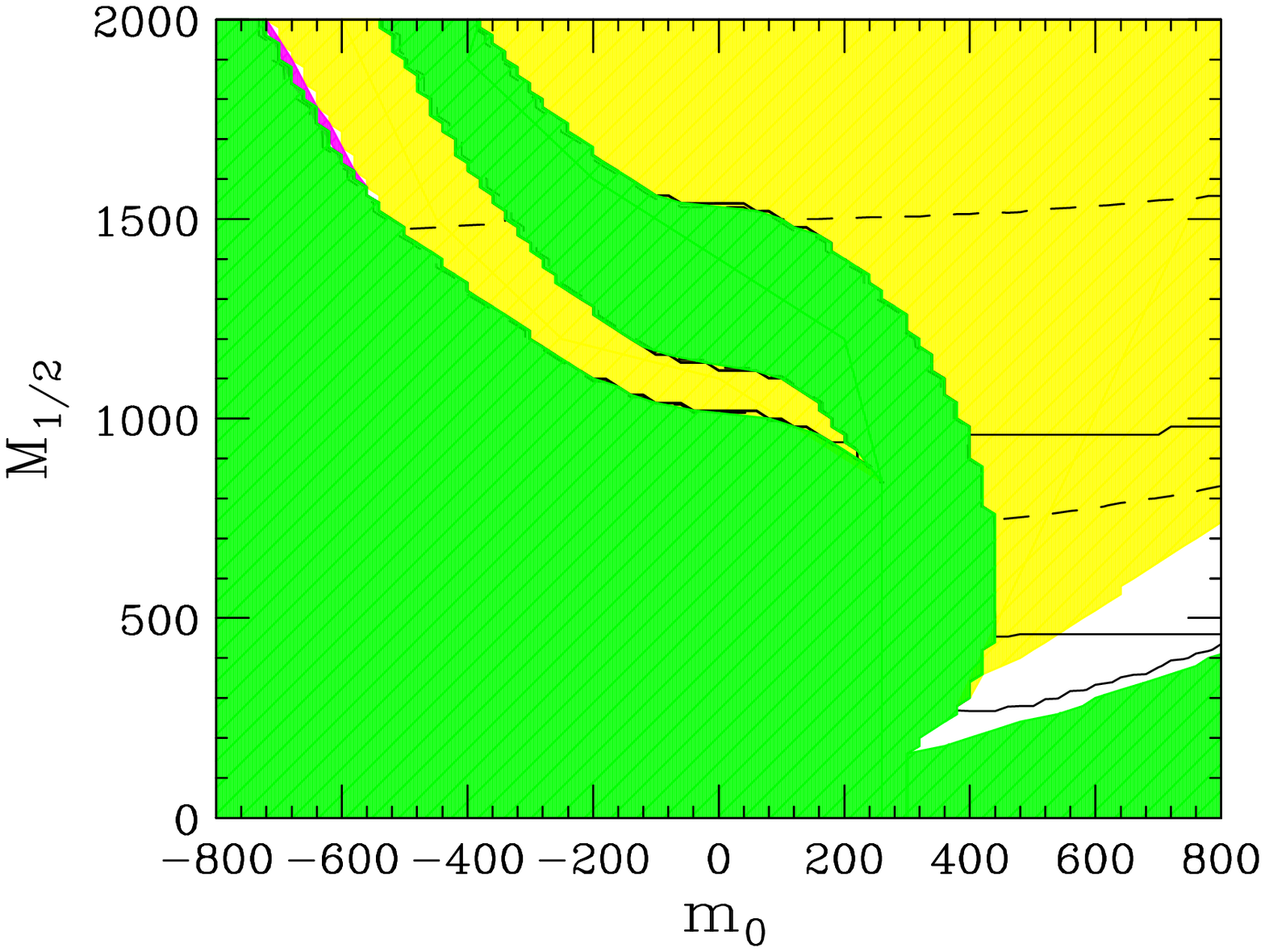}
} \caption{Higgs boson masses for mSUGRA extended to $m_0<0$ for $A_0 =
0$, $\mu > 0$, and $\tan\beta = 10$~(left) and $\tan\beta =
60$~(right).  The SM-like Higgs boson $h^0$ mass contours~(solid) have
values $114~\gev$, $118~\gev$, and $122~\gev$ from bottom to top.  The
$A^0$ mass contours~(dashed) have values $500~\gev$, $1~\tev$, and
$1.5~\tev$ from bottom to top.  The heavy CP-even and charged Higgs
scalars are approximately degenerate with the $A^0$.}
\label{fig:higgs}
\end{figure}

Last, Higgs boson masses are given in \figref{higgs}.  The masses of
the Higgs bosons, $h^0$, $A^0$, $H^0$, and $H^\pm$ are all
insensitive to $m_0$ for $m_0<0$.  The predicted value of $m_{h^0}$
is above 114 GeV throughout the allowed $m_0<0$ region, and so
consistent with current bounds.  It increases with increasing
$M_{1/2}$, rising to approximately 122 GeV for $M_{1/2} = 1~\tev$.
$A^0$, $H^0$, and $H^\pm$ are all approximately degenerate for
$m_0<0$.

\section{Precision Experimental Constraints}
\label{sec:constraints}

We now consider constraints on these models from precision
experimental data. We focus on three processes that are well-known to
have significant sensitivity to supersymmetric contributions: the
anomalous magnetic moment of the muon $a_{\mu}$ and the rare decays
$b \to s \gamma$ and $B^0_s\to\mu^+ \mu^-$. These contributions have
been calculated using the software package micrOMEGAs,
v1.3.6~\cite{Belanger:2001fz}.

\subsection{Anomalous Magnetic Moment of the Muon}

Determining the SM value of the anomalous magnetic moment of the muon
is not straightforward because of hadronic contributions to higher
order loop processes.  These hadronic loop contributions are usually
estimated from measurements of $e^+e^-$ $\to$ hadrons or $\tau$ $\to$
hadrons. The resulting SM predictions for $a_{\mu}$ are $a_{\mu} =
116\, 592\, 018\,(63)\,\times \, 10^{-11}$ if the $\tau$ data are
used, and $a_{\mu} = 116\,591\,835\,(69)\,\times\, 10^{-11}$ if the
$e^+e^-$ data are used~\cite{Davier:2003pw,Hagiwara:2003da}.  Given
theoretical assumptions required to use the $\tau$ data, the $e^+e^-$
value is generally judged to be more reliable~\cite{Passera:2005mx}.
These should be compared to the measured value $a_{\mu} = 116\, 592\,
080\,(60)\,\times\, 10^{-11}$ from the Muon $(g-2)$
Collaboration~\cite{Bennett:2002jb,Bennett:2004pv}.  Taking the SM
value for $a_{\mu}$ using the $e^+e^-$ data, there is a discrepancy
between theory and experiment of $\delta a_\mu= 245 \times 10^{-11}$,
a deviation of approximately $3 \sigma$.

\begin{figure}
\resizebox{6.0in}{!}{
\includegraphics{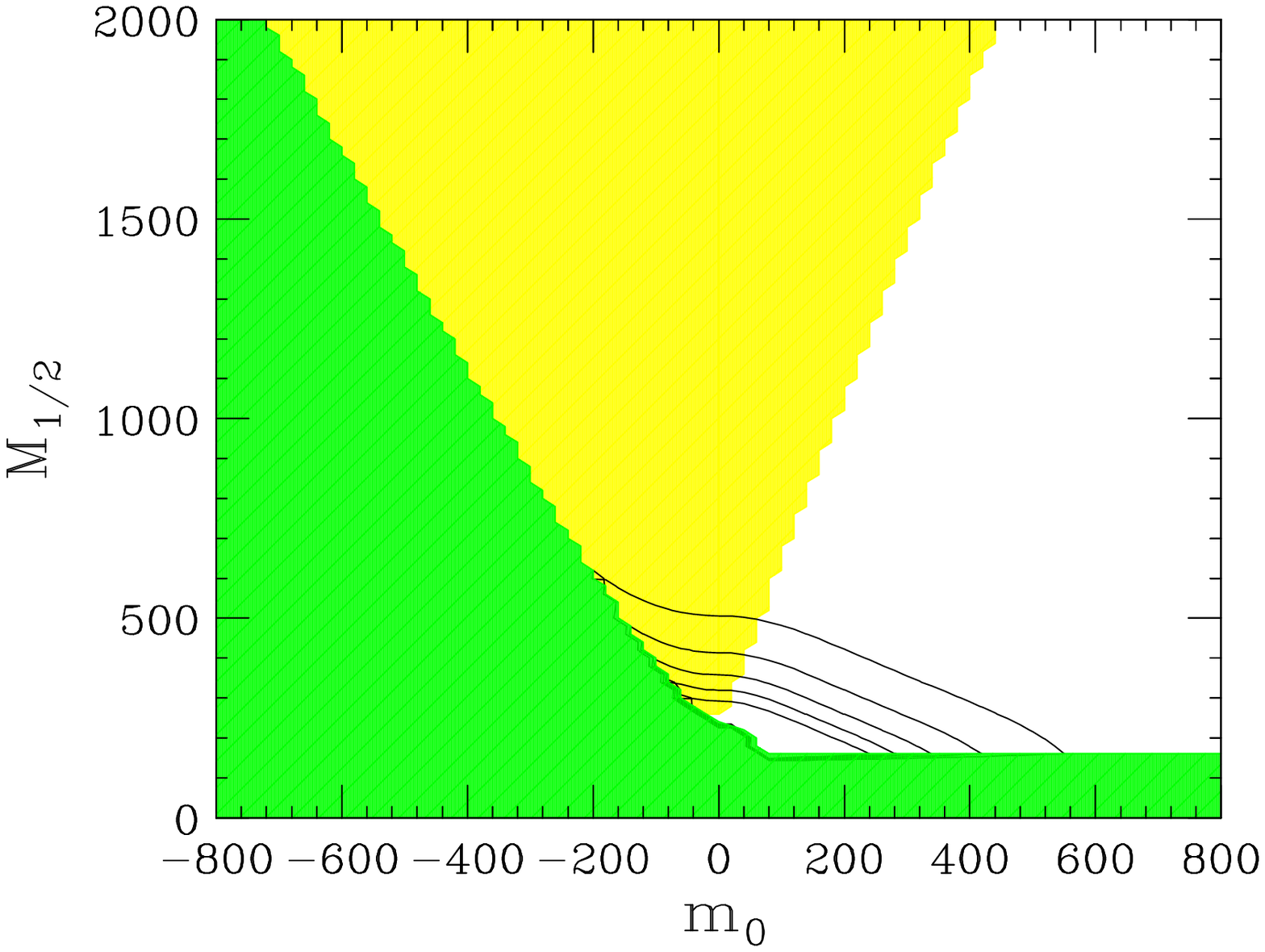} \qquad
\includegraphics{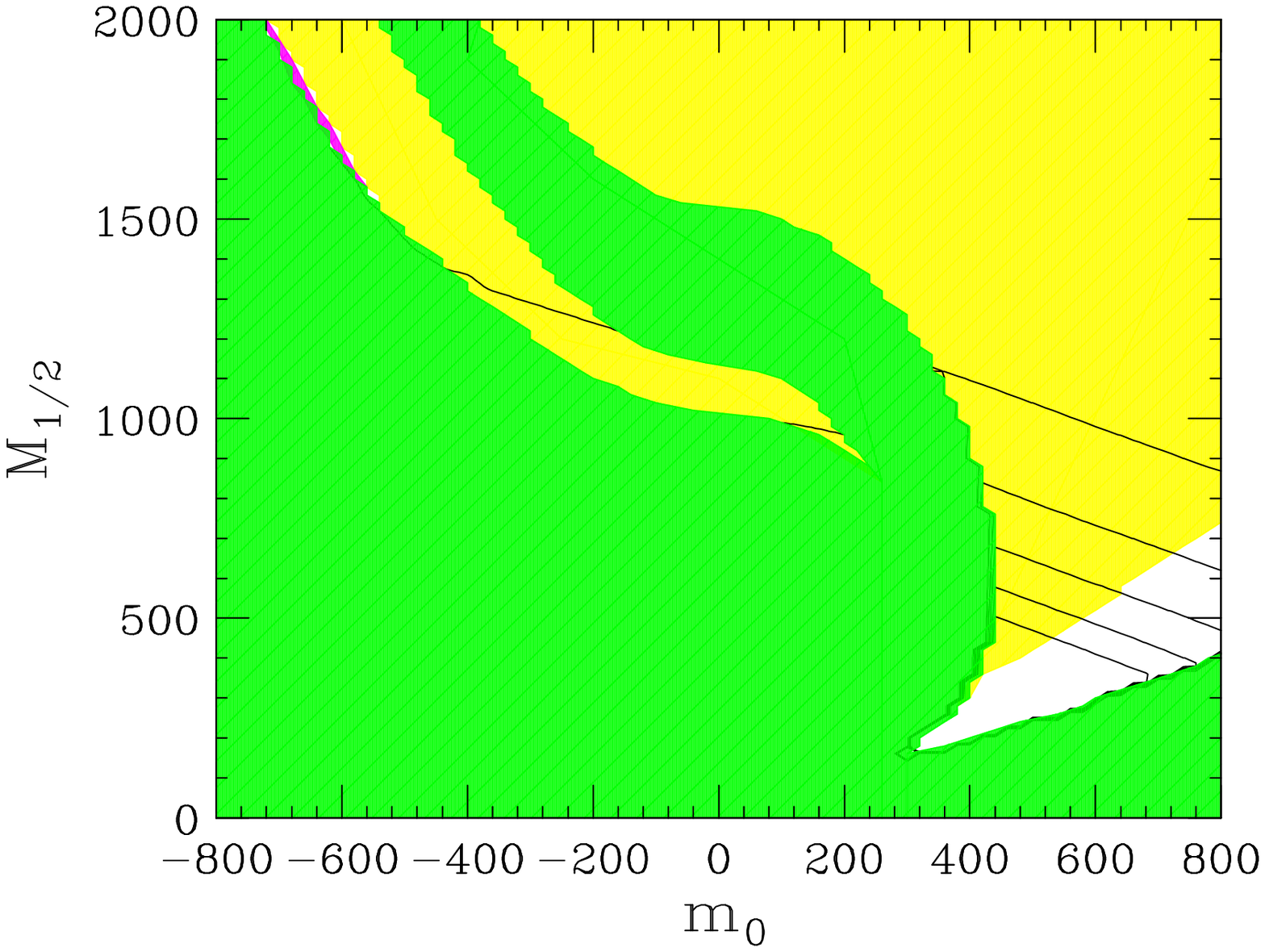}
} \caption{The SUSY contribution to $a_{\mu}$ for mSUGRA extended to
$m_0<0$ for $A_0 = 0$, $\mu > 0$, and $\tan\beta = 10$~(left) and
$\tan\beta = 60$~(right).  {}From bottom to top the contour values
are 300, 250, 200, 150, and 100 in units of $10^{-11}$.
\label{fig:gm2} }
\end{figure}

The SUSY contribution to $a_{\mu}$ in mSUGRA with $m_0<0$ is shown in
\figref{gm2}. A deviation consistent with the discrepancy between
experiment and the $e^+e^-$ SM prediction may be obtained for $\mu>0$
and light neutralinos and sleptons or light charginos and sneutrinos.
As noted in \secref{mass}, sleptons are light along the entire $m_0 <
0$ experimentally excluded border, but the gauginos increase in mass
as $M_{1/2}$ increases. A large SUSY contribution to $a_{\mu}$ is
therefore found only for relatively small $M_{1/2}$.  The $3\sigma$
deviation mentioned above may be explained, for example, for
$\tan\beta = 10$ and $M_{1/2} \sim 300~\gev$.

\subsection{\boldmath{$B(b \to s \gamma)$}}

\begin{figure}
\resizebox{6.0in}{!}{
\includegraphics{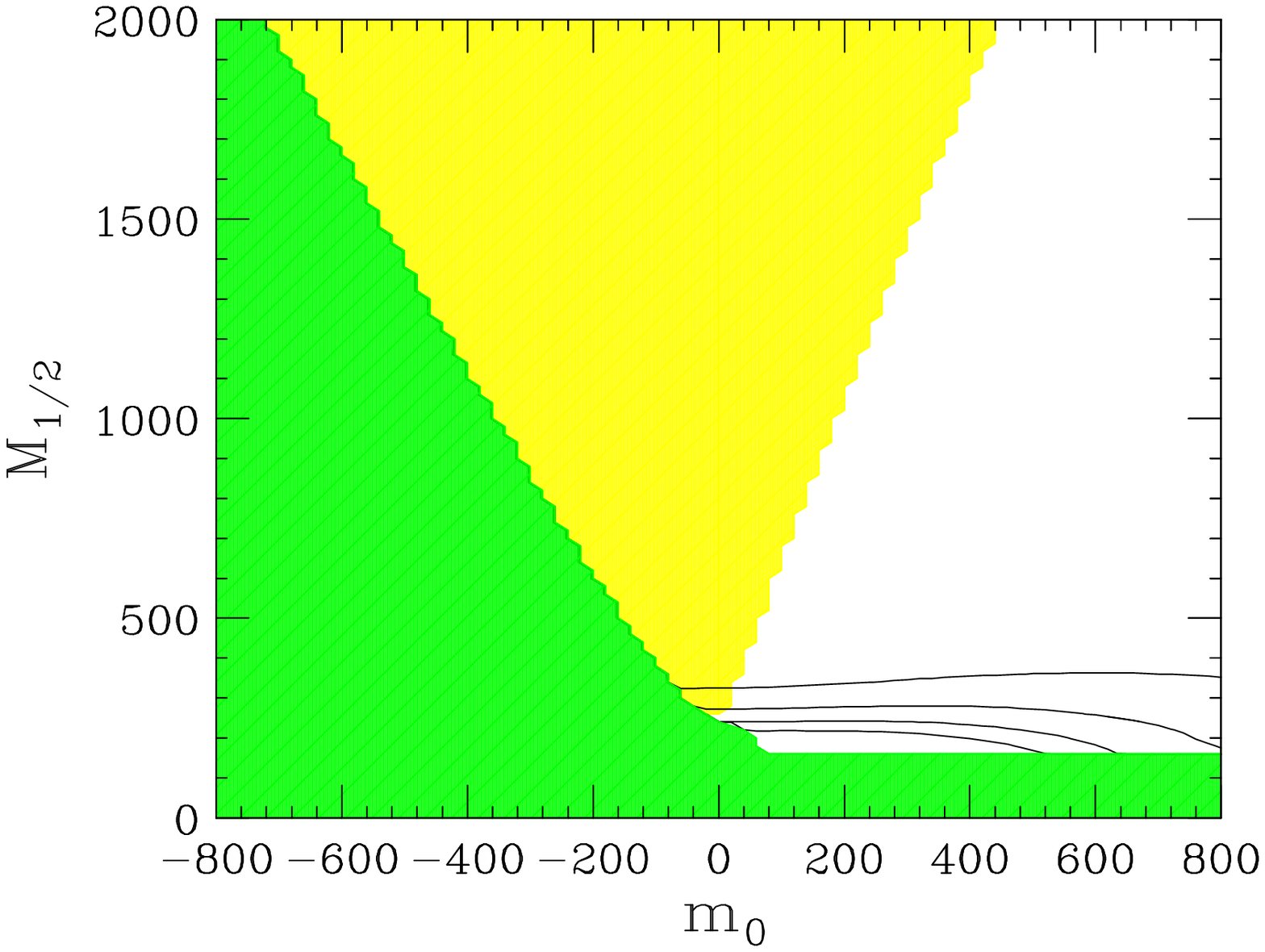} \qquad
\includegraphics{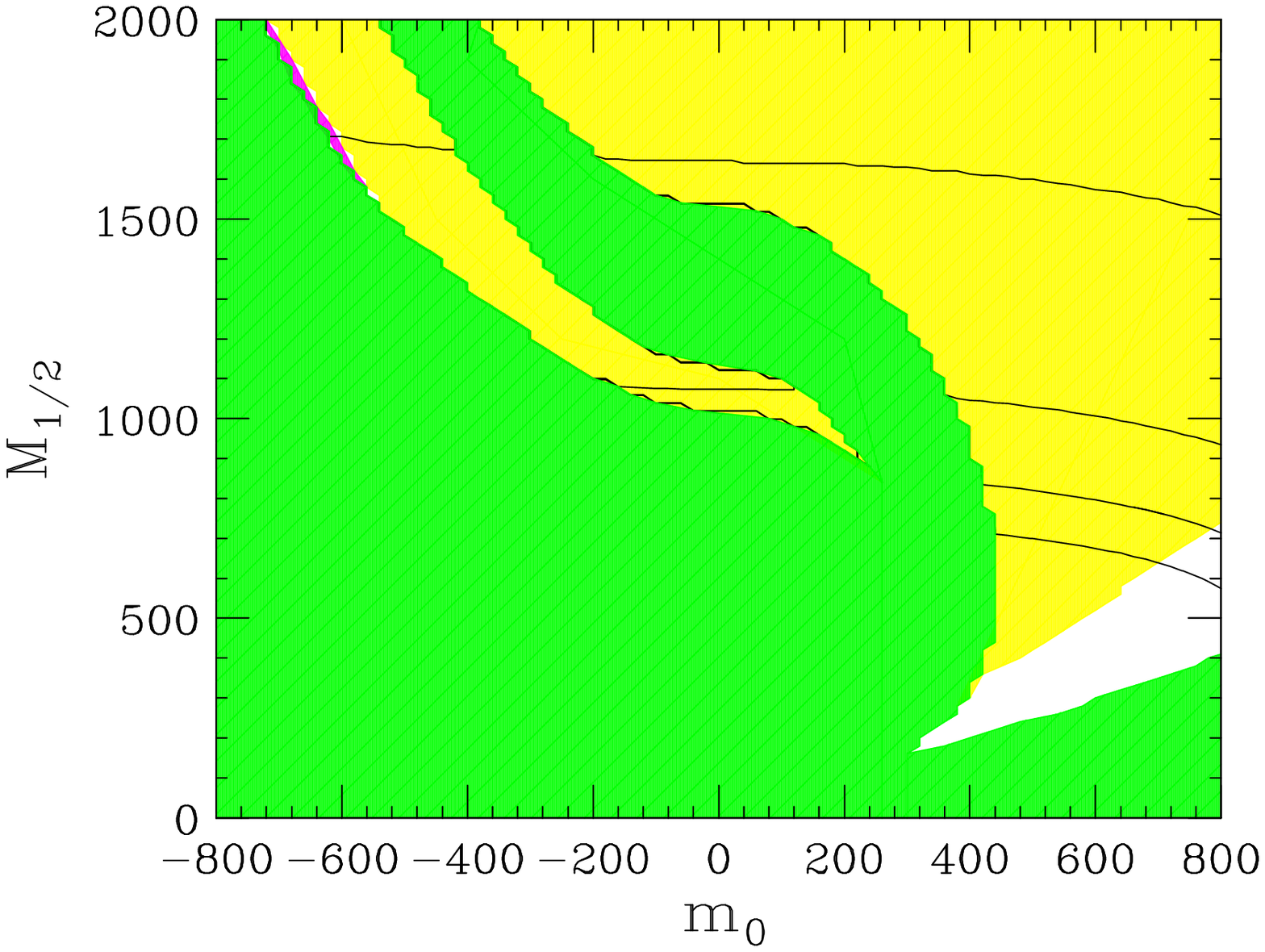}
} \caption{$B(b\to s\gamma)$ for mSUGRA extended to $m_0<0$ for $A_0
= 0$, $\mu > 0$, and $\tan\beta = 10$~(left) and $\tan\beta =
60$~(right).  The values from bottom to top are 3.10, 3.25, 3.40, and
3.55 in units of $10^{-4}$.  In the $\tan\beta = 10$ panel, $B(b \to
s \gamma)$ does not exceed $3.59 \times 10^{-4}$. \label{fig:bsgamma}
}
\end{figure}

The flavor changing neutral current transition $b \to s \gamma$ has a
branching fraction measured to be
\begin{equation}
B(b\to s\gamma) = \left \{
\begin{array}{l}
3.21\pm0.43\pm0.27^{+0.18}_{-0.10}\times 10^{-4}
~\text{(CLEO)}~\mbox{\protect\cite{Chen:2001fj}} \\
3.88\pm0.36\pm0.37^{+0.43}_{-0.23}\times 10^{-4}
~\text{(BABAR)}~\mbox{\protect\cite{Aubert:2002pd}} \\
3.55\pm0.32\pm0.30^{+0.11}_{-0.07}\times 10^{-4}
~\text{(BELLE)}~\mbox{\protect\cite{Koppenburg:2004fz}}
\end{array}
\right. \ . \label{bsg}
\end{equation}
The SM prediction is $3.79^{+0.36}_{-0.53} \times
10^{-4}$~\cite{Gambino:2001ew}.  The $m_0<0$ mSUGRA predictions for
$B(b \to s \gamma)$ are shown in \figref{bsgamma}.  The experimental
and SM theory values agree within errors, and so $b \to s \gamma$
may, in principle, eliminate models with light squarks and gauginos.
As can be seen in \figref{bsgamma}, however, supersymmetric effects
in the plotted $\m_0<0$ region are never large enough to create a
discrepancy between these mSUGRA models and experiment.

\subsection{\boldmath{$B(B^0_s \to \mu^+\mu^-)$}}

\begin{figure}
\resizebox{3.1in}{!}{
\includegraphics{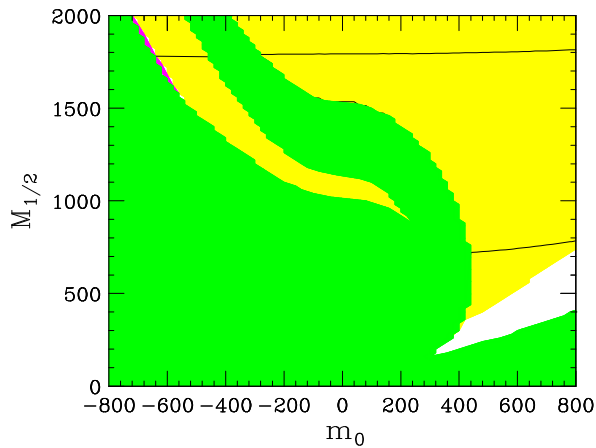}
} \caption{$B(B^0_s \to \mu^+ \mu^-)$ for mSUGRA extended to $m_0<0$
for $A_0 = 0$, $\mu > 0$, and $\tan\beta = 60$. The lower contour is
CDF II's experimental upper bound $1.5 \times 10^{-7}$, and the upper
contour is for $B(B^0_s \to \mu^+ \mu^-) = 1.0\times 10^{-8}$.
\label{fig:bsmumu} }
\end{figure}

The branching fraction for $B^0_s$ decaying to two leptons is an
important measurement for constraining supersymmetric models with
large $\tan\beta$~\cite{Choudhury:1998ze,Babu:1999hn,%
Chankowski:2000ng,Bobeth:2001sq,Huang:2002ni}.  The decay is enhanced
by $(\tan\beta)^6$ for large $\tan\beta$. The current experimental
bound is $B(B^0_s \to \mu^+ \mu^-) < 1.5 \times 10^{-7}$ from CDF II,
based on $364~\text{pb}^{-1}$ of data~\cite{Abulencia:2005pw}, while
the SM prediction is $3.42 (54) \times
10^{-9}$~\cite{Buchalla:1993bv}.  For small $\tan\beta$, mSUGRA with
$m_0 < 0$, along with other SUSY models, predicts deviations far below
current experimental bounds. These deviations will not be probed until
the LHC.  However, for large $\tan\beta$, observable deviations are
predicted even in Tevatron data.  As shown in \figref{bsmumu}, the
current Tevatron data do not exclude additional parameter space.
Nevertheless, future improvements to sensitivities of $\sim 10^{-8}$
will probe the $m_0<0$ region all the way up to $M_{1/2} \sim
1.8~\tev$ for $\tan\beta = 60$, and will also be sensitive to models
with more moderate values of $\tan\beta$.

\section{Collider Signals}
\label{sec:collider}

\begin{figure}
\resizebox{6.0in}{!}{
\includegraphics{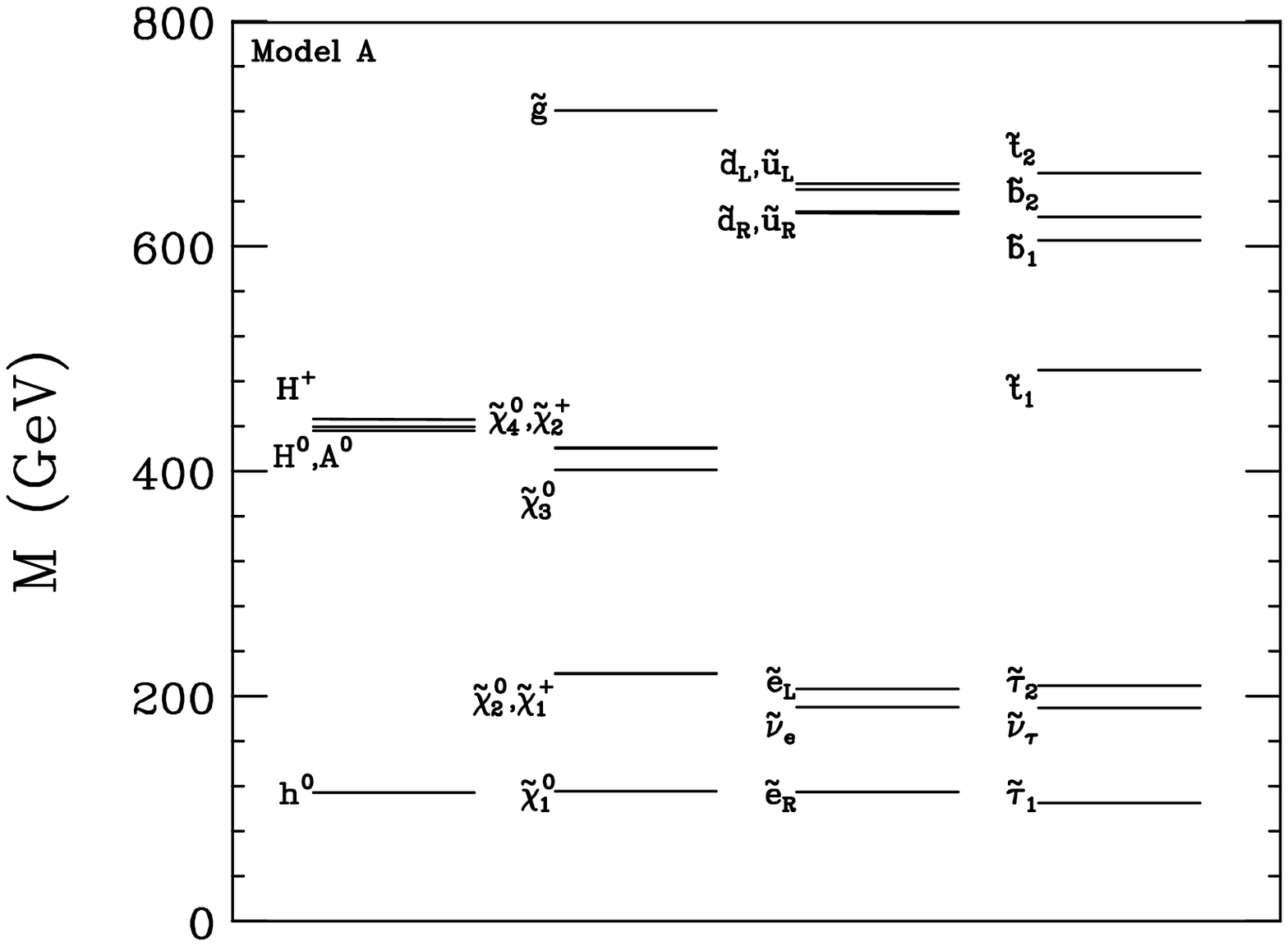} \qquad
\includegraphics{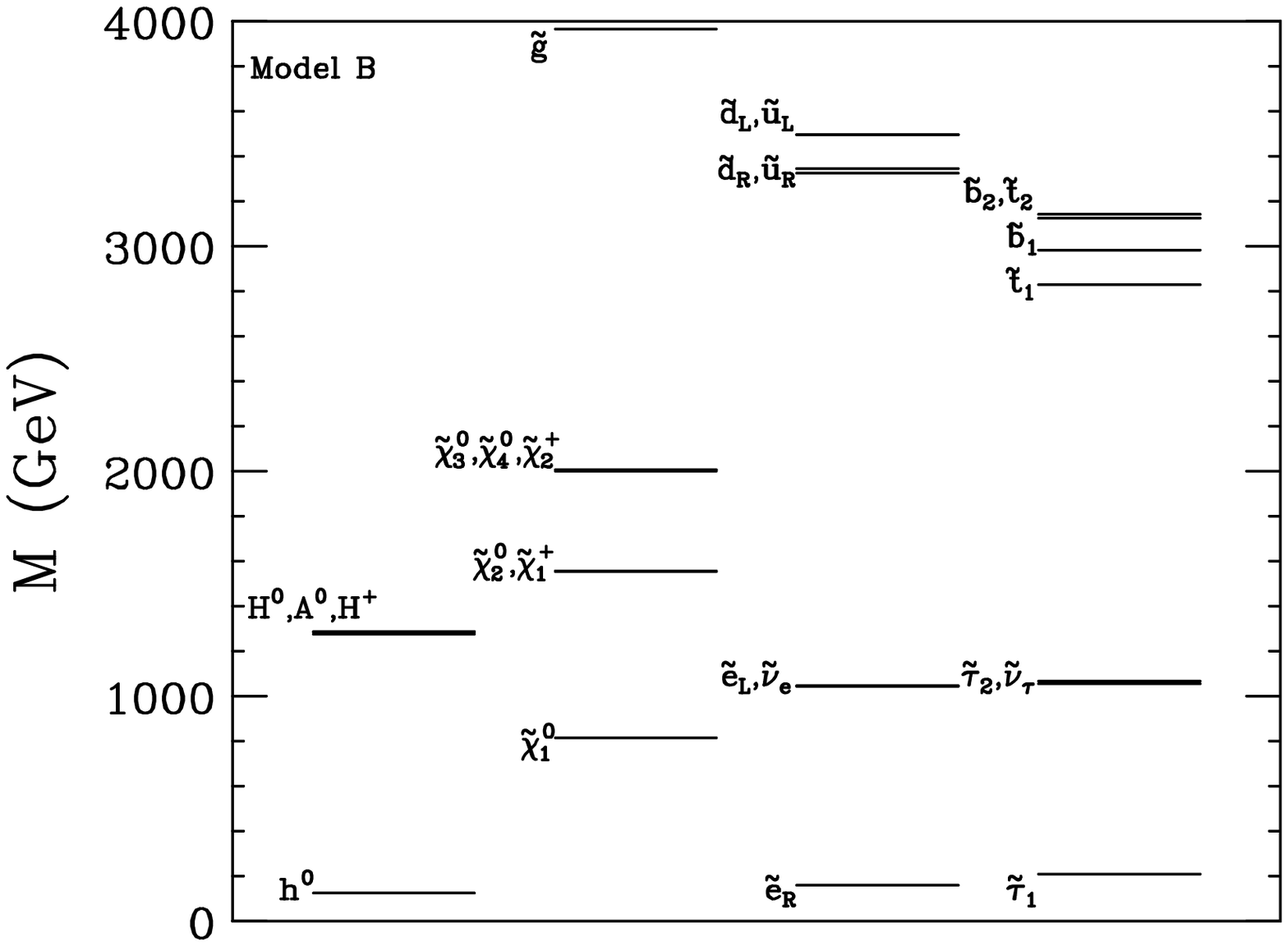}
} 
\caption{Superpartner spectra for Model A, the stau NLSP point $m_0 =
  -40~\gev$, $M_{1/2} = 300~\gev$, $\tan\beta = 10$ (left) and for
  Model B, the selectron NLSP point $m_0 = -700~\gev$, $M_{1/2} =
  1900~\gev$, $\tan\beta$ = 60 (right).  In both cases, $A_0 = 0$ and
  $\mu > 0$.  These models correspond to the parameter points
  highlighted with $\oplus$ symbols in \protect\figref{lsp}; the RGEs
  of their scalars are displayed in \protect\figref{rgelepton}, and
  their weak-scale parameters are given in \tableref{parameters}.
\label{fig:spectra} }
\end{figure}

As is well-known, the LHC will provide an extremely powerful probe of
weak-scale supersymmetry in the next few years.  Here we discuss the
implications of $m_0<0$ mSUGRA for the LHC and the proposed ILC.

The prototypical signature of ($R$-parity conserving) supersymmetry at
hadron colliders is missing transverse energy.  In the case of mSUGRA
with $m_0<0$, however, all SUSY events result in the production of two
metastable charged sleptons. These have lifetimes of seconds to
months, and so pass through collider detectors without decaying.
These models therefore provide a conventional setting for what might
otherwise be considered to be rather exotic signals, such as highly
ionizing tracks and time-of-flight
signatures~\cite{Drees:1990yw,Goity:1993ih,Nisati:1997gb,Feng:1997zr}.
Even a few events will provide unmistakable signals.

As examples, let us consider the benchmark models indicated in
\figref{lsp}.  The RGEs for scalars in these models were shown in
\figref{rgelepton}. In \figref{spectra}, we display the full
superpartner spectrum for each of these models, and in
\tableref{parameters} we list all mass parameters, which define these
models at the electroweak scale.

\begin{table}[tb]
\begin{tabular}{|c|c|c|c|}
\hline ISAJET Specification & Parameter
& \quad Model A \quad 
& \quad Model B \quad
\\ \hline
MSSMA  & $m_{\tilde{g}}$ \quad $\mu$ & 720.70 395.95 & 3964.95 1994.33 \\
       & $m_{A^0}$ $\tan\beta$  & 436.09   10.00 & 1274.81 60.00 \\ \hline
MSSMB  & $m_{\tilde{q}_1}$ $m_{\tilde{d}_R}$ $m_{\tilde{u}_R}$ &
652.21 601.12 603.80 & 3352.61 3178.77 3198.46 \\
       & $m_{\tilde{l}_1}$ $m_{\tilde{e}_R}$ & 199.04 103.81 & 
1027.84   67.57 \\ \hline 
MSSMC &$m_{\tilde{q}_3}$ $m_{\tilde{b}_R}$ $m_{\tilde{t}_R}$ &
578.76  598.21  502.38 & 2987.96 2860.45 2727.32 \\
      &$m_{\tilde{l}_3}$ $m_{\tilde{\tau}_R}$& 198.49 101.56 & 1040.40
255.73 \\
      &$A_t$ $A_b$ $A_{\tau}$ & 548.08 805.29 186.50 & 2788.73 3147.17
14.15 \\ \hline 
MSSMD &$m_{\tilde{q}_2}$ $m_{\tilde{s}_R}$ $m_{\tilde{c}_R}$ 
 & Same as MSSMB (default) & Same as MSSMB (default) \\
 & $m_{\tilde{l}_2}$ $m_{\tilde{\mu}_R}$& & \\ \hline
MSSME & $M_1$ $M_2$ & 120.17  231.47 & 831.43 1527.66 \\ \hline
& $ \delta \text{a}_{\mu}$ & $296 \times 10^{-11}$ & $78.1 \times 10^{-11}$ \\ \hline
& $B\left(b \rightarrow s\gamma \right)$ & $3.49 \times 10^{-4}$  &$3.57
\times 10^{-4}$ \\
\hline
& $B\left(B^{0}_{s} \rightarrow \mu^+ \mu^- \right)$ & $3.21 \times
10^{-9}$ & $8.84 \times 10^{-9}$ \\ \hline
\end{tabular}
\caption{Mass parameters in GeV and the predicted values for precision
  observables for benchmark Models A and B.  The masses in category
  MSSMA are physical masses; all other masses listed are soft
  SUSY-breaking parameters specified at the electroweak scale. }
\label{table:parameters}
\end{table}

In Model A, the stau NLSP model of the right panel in
\figref{spectra}, all superpartners have masses under 1 TeV.  This
model is an excellent benchmark model.  It explains the $3\sigma$
deviation in $a_{\mu}$, and preserves the agreement between the SM and
experimental values of $B(b \to s \gamma)$.  In addition, electroweak
symmetry is broken radiatively and naturally, with $\mu \sim
400~\gev$.

This stau NLSP model has a total SUSY cross section at the LHC of
$\sigma_{\text{LHC}}(14~\tev) = 13.6~\pb$, as determined by
ISAJET~\cite{Paige:2003mg}.  Even with an integrated luminosity of
$100~\pb^{-1}$, this implies over 1000 SUSY events, each with two
metastable sleptons.  In many of these, the sleptons will be slow
enough to be seen as highly-ionizing tracks, providing a spectacular
signal of new physics in even the first year of LHC operation.  With
more luminosity, large numbers of sleptons may be collected and their
decays studied, making possible a variety of measurements with
implications for cosmology, astrophysics and
supergravity~\cite{Feng:2003xh,Feng:2004zu,Buchmuller:2004rq,%
Feng:2004gn,Hamaguchi:2004df,Feng:2004yi,DeRoeck:2005bw,%
Kaplinghat:2005sy,Cembranos:2005us,Jedamzik:2005sx,Sigurdson:2003vy,%
Profumo:2004qt,Roszkowski:2004jd,Cerdeno:2005eu,Jedamzik:2004er,%
Kawasaki:2004qu,Ellis:2005ii,Lamon:2005jc}.

Model B, the selectron NLSP model indicated in \figref{lsp}, is a
complementary benchmark model. As seen in \figref{spectra}, the model
has squarks and gluinos around $3-4~\tev$, neutralinos, sneutrinos,
non-SM type Higgs bosons, and left-handed sleptons around $1-2~\tev$,
a $210~\gev$ stau, $160~\gev$ selectron and smuon, and finally a
$124~\gev$ Higgs boson.  The squarks and gluinos are too heavy to be
produced with large cross section at the LHC. The biggest sources of
SUSY particles at the LHC are direct Drell-Yan production of NLSP
pairs, leading to 2 metastable charged sleptons, and Drell-Yan
production of the heavier sleptons, leading to even more unusual
events with 2 taus, 2 muons/electrons, and 2 metastable charged
sleptons.  The total SUSY production cross section for this model is
$\sigma_{\text{LHC}}(14~\tev) = 41~\fb$.  This signal will be
challenging to find in the first year of LHC running, but at the
target luminosity of $100~\ifb/\yr$, the LHC will produce 4100 SUSY
events per year.

At the ILC, the total SUSY production cross sections are
$\sigma_{\text{ILC}}(500~\gev) = 1.35~\pb$ for the stau NLSP Model A
and $\sigma_{\text{ILC}}(500~\gev) = 137~\fb$ for the selectron NLSP
Model B.  Although squarks and gluinos are out of reach, the ILC will
produce large numbers of sleptons.  All the usual advantages of the
ILC will allow detailed studies of the SUSY parameter space. In
addition, the ability to produce sleptons at low velocities implies
that they may be more easily trapped and studied than at the LHC.

\section{Conclusions}
\label{sec:conclusions}

In this study, we have extended the well-known framework of mSUGRA to
$m_0<0$.  To our knowledge, this part of parameter space has not been
considered previously, perhaps because it contains a charged slepton
as the lightest SM superpartner.  If the gravitino is the LSP,
however, cosmological difficulties with CHAMPs are avoided, and this
extended parameter space is allowed.  We have noted that it is
consistent with all limits from direct searches and constraints from
low-energy precision measurements.  In addition, we find models with
qualitatively novel mSUGRA superpartner spectra, which predict
spectacular signals with metastable charged sleptons at future
colliders.

Some of the cosmology of these models will be considered in a future
work.  It would be interesting to extend this work to more general
models.  For example, as argued above, we expect that the selectron
may emerge as the NLSP generically in models with non-unified Higgs
masses, and there may well be other interesting phenomena.  It would
also be worthwhile to determine the reach of the LHC for various
luminosities in the $m_0<0$ parameter space; given how spectacular the
signal of metastable charged particles will be, these models provide a
welcome example in which supersymmetry may be discovered and studied
in even the first year of LHC running.

\begin{acknowledgments}
The work of JLF is supported in part by NSF CAREER grant
No.~PHY--0239817, NASA Grant No.~NNG05GG44G, and the Alfred P.~Sloan
Foundation. AR is supported in part by NSF Grant No.~PHY--0354993.
\end{acknowledgments}


\end{document}